\documentclass[useAMS,usenatbib]{mn2e}

\usepackage{amsmath, amssymb}
\usepackage{epsfig}         
\usepackage{graphicx, float}
\usepackage{multirow}

\usepackage{natbib}
\usepackage{aas_macros}

\usepackage{url}
\usepackage{hyperref}

\usepackage{color}
\definecolor{oldcolor}{rgb}{.3,.3,.3}

\newcommand{\MCn}[1]{#1} 

\def\Mpc{~{\rm Mpc}}
\def\Mpch{~h^{-1} {\rm Mpc}}

\def\kpc{~\rm kpc}

\def\Msun{\rm{M}_{\odot}}
\def\Msolar{~h^{-1} \rm{M}_{\odot}}
\def\kms{~\rm{km/s}}

\newcommand{\rockstar}{\textsc{rockstar}}
\newcommand{\subfind}{\textsc{subfind}}

\newcommand{\MI}{\textsc{MS}}
\newcommand{\MII}{\textsc{MS-II}}
\newcommand{\dove}{\textsc{WMAP7}}

\newcommand{\LR}{\rmn{LR}}
\newcommand{\HR}{\rmn{HR}}

\newcommand {\lcdm}{$\Lambda$CDM}

\newcommand{\g}{{>}} 
\newcommand{\cum}{({>}\nu)}  
\newcommand{\Ncum}{\overline{N}({>}\nu)}  
\newcommand{\Nnu}{\overline{N}(\nu)}  

\newcommand{\gsim}{\raisebox{-0.3ex}{\mbox{$\stackrel{>}{_\sim} \,$}}}
\newcommand{\lsim}{\raisebox{-0.3ex}{\mbox{$\stackrel{<}{_\sim} \,$}}}
\newcommand{\eq}[1]{Eqn.~\eqref{#1}}

\newcommand{\refsec}[1]{Section~\ref{#1}}
\newcommand{\refappendix}[1]{Appendix~\ref{#1}}
\newcommand{\reftab}[1]{Table~\ref{#1}}
\newcommand{\reffig}[1]{Fig.~\ref{#1}}

\newcommand{\figDir}{fig_pdf/}

\newcommand{\wang}{Wang12}
\newcommand{\bkOne}{BK10}

\title[Subhalo statistics of galactic halos: beyond the resolution limit]
{Subhalo statistics of galactic halos: beyond the resolution limit}

\author[Cautun~et~al.]
{\parbox{\textwidth}{       
	    Marius Cautun$^{1,2}$\thanks{E-mail : m.c.cautun@durham.ac.uk},
        Wojciech A. Hellwing$^{1,3}$\thanks{E-mail : pchela@icm.edu.pl},
        Rien van de Weygaert$^{2}$, 
        Carlos S. Frenk$^{1}$,
        Bernard~J.~T.~Jones$^{2}$
        and Till Sawala$^{1}$\vspace{.4cm}} \\
$^1$   Department of Physics, Institute for Computational Cosmology, University of Durham, South Road Durham DH1 3LE \\
$^2$   Kapteyn Astronomical Institute, University of Groningen, P.O. Box 800, 9747 AV Groningen, The Netherlands \\
$^3$   Interdisciplinary Centre for Mathematical and Computational Modelling, University of Warsaw, ul. Pawi\'nskiego 5a, Warsaw, Poland\\
}

\begin{document}


\maketitle

\begin{abstract}
  We study the substructure population of Milky Way (MW)-mass halos
  in the \lcdm{} cosmology using a novel procedure to extrapolate
  subhalo number statistics beyond the resolution limit of N-body
  simulations. The technique recovers the mean and the variance of the
  subhalo abundance, but not its spatial distribution.  It extends the
  dynamic range over which precise statistical predictions can be made
  by the equivalent of performing a simulation with 50 times higher
  resolution, at no additional computational cost. We apply this
  technique to MW-mass halos, but it can easily be applied to halos
  of any mass. We find up to $20\%$ more substructures in MW-mass
  halos than found in previous studies. Our analysis lowers the mass
  of the MW halo required to accommodate the observation that the MW
  has only three satellites with a maximum circular velocity
  $V_\rmn{max}\ge30\kms$ in the \lcdm{} cosmology.  The probability of
  having a subhalo population similar to that in the MW is $20\%$ for
  a virial mass, $M_{200}=1\times10^{12}\Msun$ and practically zero
  for halos more massive than $M_{200}=2\times10^{12}\Msun$.
\end{abstract}



\section{Introduction}
\label{sec:introduction}

The standard `$\Lambda$ cold dark matter' (\lcdm{}) cosmological model
has been found to give a good description of structure formation
and evolution on scales $\gsim10\Mpc$. This has been confirmed by
multiple observational probes: the cosmic microwave background
temperature anisotropies \citep[eg.][]{Komatsu2011,Planck2013_XVI},
large-scale galaxy clustering \citep[eg.][]{Cole2005} and the
expansion history of the Universe
\citep[eg.][]{Clocchiatti2006,Guy2010}. On smaller scales, the \lcdm{}
predictions are more difficult to extract and test due both to the
non-linear evolution of the matter distribution and the complex
hydrodynamical processes that drive galaxy formation and evolution.
Nonetheless, it is this regime that is especially interesting and
important for cosmology as it can potentially constrain the nature of
the dark matter and the baryonic processes involved in galaxy
formation. Our own MW galaxy and its satellites play a
crucial role in this due to their proximity which enables in-depth
studies.

Several of the apparent points of tension between observations and
\lcdm{} predictions are seen in the properties of the MW and its
satellites. The phrase ``missing satellites problem" is often
incorrectly used to refer to the apparent discrepancy between the
large number of dark matter subhalos in N-body simulations, first
highlighted by \cite{Moore1998}, and the handful of satellites
detected around the MW. In fact, this ``problem'' simply reflects the
well-known fact that most of the dark matter subhalos never manage to
acquire a visible galaxy because of inevitable physical processes,
such as reionization and the injection of supernova energy, that are
an intrinsic part of galaxy formation
\citep{Bullock2000,Benson2002,Somerville2002}.

A more significant ``satellite problem,'' recognized as such already
by \cite{Klypin1999} and \cite{Moore1999}, is the apparent discrepancy
between the distribution of the maximum circular velocities of the
most massive subhalos in \lcdm{} simulations and the inferred values
for the MW's satellites.  Various arguments based on the kinematics of
the nine bright ``classical'' dwarf spheroidal satellites of the MW
suggest that their subhalos have maximum circular velocities
$V_\rmn{max}\lsim30\kms$
\citep{Penarrubia2008,Strigari2008,Lokas2009,Walker2009,Wolf2010,Strigari2010,Boylan-Kolchin2011a,Boylan-Kolchin2012a}.
These are lower than the values for the most massive subhalos in
simulations of galactic halos such as the high-resolution simulations
of the Aquarius project \citep{aquarius2008}.  Specifically,
\cite{Boylan-Kolchin2011a,Boylan-Kolchin2012a} brought attention to
the observation that these simulations typically produce around eight
subhalos with $V_\rmn{max}>30\kms$, whereas in the MW only the two
Magellanic Clouds and the Sagittarius dwarf are thought to reside in
subhalos with such high circular velocities. This raises the
possibility that there could be several massive substructures in the
MW without a luminous galaxy in them. The high mass of these
subhalos, however, makes this rather unlikely given that less massive
subhalos do have satellite galaxies associated with them.

A possible solution to this so-called ``too-big-to-fail'' problem was
proposed by \cite{Wang2012} (hereafter \wang{}) who showed that the
presence of only three massive satellites in our galaxy is consistent
with \lcdm{} predictions provided the mass of the MW dark halo is
${\sim}1\times10^{12}\Msun$, around half the average mass of the halos
in the Aquarius simulations analyzed by
\cite{Boylan-Kolchin2011a,Boylan-Kolchin2012a} \citep[see
also][]{Purcell2012,Vera-Ciro2013}.  \wang{} used the invariance of the scaled
subhalo velocity function
\citep[e.g.][]{Moore1999,Kravtsov2004,Zheng2005,aquarius2008,Weinberg2008}
to extend the subhalo number statistics derived from N-body
simulations of large cosmological volumes to galactic halos.  This
allowed them to compute, as a function of halo mass, the probability
of having a satellite population similar to that of the MW. The
outcome of this calculation favours a MW halo mass at the lower end of
the range spanned by recent estimates
\citep{Wilkinson1999,Sakamoto2003,Battaglia2005,Dehnen2006,Smith2007,Li2008,Xue2008,Gnedin2010,Guo2010,Watkins2010,Busha2011b,Piffl2014}.

Characterising how typical the MW satellites are in \lcdm{} requires
large samples of simulated MW-mass halos. Simulations of large
cosmological volumes provide these but, so far, only at relatively low
resolution, probing only the most massive subhalos ($\lsim10$
substructures per MW halo; \citealt{millSim2,Klypin2011}). By
contrast, high-resolution ``zoom'' simulations of individual MW-like
halos resolve substructures down to much lower masses, but because of
their large computational cost, only a few examples have been
simulated so far and these are not guaranteed to be characteristic of
a MW-like halo population
\citep{Diemand2008,aquarius2008,Stadel2009}. Some of the alleged
points of tension between observations and models rely on such
high-resolution, but limited-sample studies, and one cannot exclude the
possibility that these discrepancies reflect the inherent cosmic
variance of small-volume studies.

In this work we introduce a new method for extending subhalo
statistics beyond the resolution limit available to cosmological
simulations. This allows us to investigate the statistical properties
of the subhalo population of a representative sample of MW-mass halos
down to substructures with $V_\rmn{max}\gsim15\kms$, which represents
a threefold increase in the $V_\rmn{max}$ range compared to related
previous studies (e.g. \citealt{Boylan-Kolchin2010}, hereafter
\bkOne{}; \wang{}). Making use of our extrapolation method, we can
check previous subhalo count results, such as those of \wang{}, over a
larger dynamical range in subhalo mass. In particular, we analyse the
dependence of the mean subhalo count on halo mass and revisit the
probability of finding a satellite population similar to that in the
MW. 

\MCn{Our extrapolation method should not be confused with semi-analytical models for DM substructure \citep[e.g.][and the later refinements of \citealt{Zentner2005a,Jiang2014}]{Benson2002a}. Our method statistically generates the correct subhalo abundance from the partial information available in a simulation of limited resolution. In contrast, semi-analytical models are based on halo merger trees and on the treatment of the various physical processes that affect the evolution of subhalos. While such models are significantly faster than numerical simulations, they are limited because of their approximate treatment of relevant physical processes.}

In \refsec{sec:data_analysis} we describe the simulations we use 
and the halo/subhalo identification algorithm. In 
\refsec{sec:resolution} we introduce the scaling
method for extending the subhalo statistics to masses that are unresolved 
in the simulations. In sections~\ref{sec:abundance} and
\ref{sec:host_mass} we investigate the subhalo population of MW-like
halos. \MCn{Given that we find significantly more subhalos than previous studies, in \refsec{sec:mw_probability} we revisit the constraints on}
the MW halo mass required to avoid the too-big-to-fail
problem. In \refsec{sec:aquarius_halos} we
study how typical the Aquarius halos are compared to a
representative sample of MW-like hosts. We end with a brief summary in
\refsec{sec:conclusion}.


\section{Data analysis}
\label{sec:data_analysis}
In this study we analyse the two high resolution Millennium
simulations\footnote{Data from the Millennium/Millennium-II simulation is available 
on a relational database accessible from 
\newline http://galaxy-catalogue.dur.ac.uk:8080/Millennium .}  
(\MI{}; \citealt{millSim} and \MII{}; \citealt{millSim2}).
Both are dark matter only simulations and make use of $2160^3$
particles to resolve structure formation in the \textit{Wilkinson
  Microwave Anisotropy Probe} (WMAP)-1 cosmogony \citep{Spergel2003}.
The \MI{} models cosmic evolution in a periodic volume of length
$500\Mpch$ with a mass per particle of $m_p=8.6\times10^8\Msolar$. The
large volume of the simulation makes it ideal for the study of
substructures in cluster and group sized objects, but it is of limited
use for MW-sized halos which are resolved with only ${\sim}10^3$
particles. The \MII{} resolves structure formation in a much smaller
box of $100\Mpch$ on a side with a particle mass of
$m_p=6.89\times10^6\Msolar$. The lower mass per dark matter particle makes it
suitable for studying MW-like halos that are resolved with around
$10^5$ particles, but its smaller volume precludes a systematic study
of higher mass objects. The parameters used in the two simulations are
given in \reftab{tab:numerical_parameters}.

\begin{table}
    \caption{The cosmological and numerical parameters of the three
      N-body simulations used in this study.} 
    \label{tab:numerical_parameters}
    \begin{tabular}{lccc}
        \hline
        Parameter & \MI{} & \MII{} & \dove{} \\
        \hline
        Box size $(h^{-1}{\rm Mpc})$    & 500 & 100 & 70.4 \\
        Particle number                 & 2160$^3$ & 2160$^3$ & 1620$^3$ \\
        Particle mass $(10^6 h^{-1}\Msun)$   & $860$ & $6.89$ & $6.2$ \\
        $\Omega_m$                      & $0.25$ & $0.25$ & $0.272$ \\
        $\Omega_\Lambda$                & $0.75$ & $0.75$ & $0.728$ \\
        $\sigma_8$                      & $0.9$  & $0.9$  & $0.81$ \\
        $h$                             & $0.73$ & $0.73$ & $0.704$ \\
        $n_s$                           & $1$    & $1$    & $0.968$ \\
        Force softening $(h^{-1}{\rm kpc})$ & $5$& $1$    & $1$ \\
        \hline
    \end{tabular}
\end{table}

The difference in the resolution of the two simulations, with equal
mass halos being resolved with 125 times more particles in \MII{}
than in \MI{}, makes it possible to carry out convergence tests and
other tests of the numerical effects on the subhalo population.

We also analyze a $1620^3$ particle N-body simulation of a volume
$70.4\Mpch$ on a side in the WMAP-7 cosmology
\citep{Komatsu2011}. This has a similar particle mass to the
\MII{}, $m_p=6.2\times10^6\Msolar$, but only a third of the \MII{}
volume. We refer to this additional simulation as \dove{} and use it
to investigate the differences between the predictions of WMAP-1 and
WMAP-7 \lcdm{} universes.

For comparative purposes we also make use of the Aquarius Project data
\citep{aquarius2008}, a set of MW-mass dark matter halos simulated at
very high resolution in the WMAP-1 cosmology. The six halos, denoted
Aq.-A through Aq.-F, were selected from the \MII{} and resimulated at
increasingly higher resolution. Here we make use of the ``level-2''
halos that have a particle mass of $\sim10^4\Msolar$ and
gravitational softening of $48\rmn{~h^{-1}pc}$. 

\subsection{Halo finder}
\label{subsec:rockstar}

We identify halos and subhalos using the \rockstar{} (Robust
Overdensity Calculation using K-Space Topologically Adaptive
Refinement) phase-space halo finder \citep{Berhoozi2011}. \rockstar{}
starts by selecting potential halos as Friends-of-Friends (FOF;
\citealt{Davis1985}) groups in position space using a large linking
length ($b = 0.28$ the mean interparticle separation). This first step
is restricted to position space to optimize the use of computational
resources, while subsequent steps employ the full 6D phase space. Each
FOF group from the first step is used to create a hierarchy of FOF
phase-space subgroups by progressively reducing the linking
length. The phase-space subgroups are selected using an adaptive
phase-space linking length such that each successive subgroup has
$70\%$ of the parent's particles. \rockstar{} uses the resulting
subgroups to identify potential halo and subhalo centres and assigns
particles to them based on their phase-space proximity. Once all
particles are assigned to halos and subhalos, an unbinding procedure
is applied to retain only gravitationally bound particles. The final
halo centres are computed from a small region around the phase-space
density maximum of each object.

The outer boundary of the halos is defined as the distance at which the
enclosed overdensity decreases below $\Delta=200$ times the critical
density, $\rho_c$. Therefore, the halo mass, $M_{200}$, and radius,
$R_{200}$, correspond to a spherical overdensity of $200\rho_c$. Using
this definition for the main halo boundary, we identify all subhalos
within distance $R_{200}$ from the host halo centre as the satellite
population. A typical MW-mass halo with $M_{200}=10^{12}\Msun$ has
$R_{200}{\approx}200\kpc$ which is smaller than the maximum distance
commonly used to identify dwarf galaxies in the MW; for example Leo I
is considered a MW satellite but it is located ${\sim}250\kpc$ from
our galaxy \citep{Karachentsev2004}. We therefore apply a second
criterion and identify as subhalos all the objects within $R_{100}$
from the host centre. The distance $R_{100}$ is the radius within
which the enclosed overdensity decreases to $100\rho_c$ and is
typically ${\sim}1.3$ times larger than $R_{200}$. We denote this
second group of subhalos as $R_{100}$ substructures.


\section{Extrapolating subhalo statistics beyond the resolution limit}
\label{sec:resolution}
There are two challenges when studying the satellite population in
numerical simulations: identifying the subhalos and correctly
determining their internal structure and orbits.  Identifying an object
made of a few tens to hundreds of particles against the background of
a much bigger halo is not trivial and most configuration-space halo
finders have difficulties finding subhalos of fewer than $50$
particles as well as larger subhalos located close to the centre of
the host. While phase-space finders (which includes \rockstar{})
perform somewhat better, they still have problems recovering the
correct properties of substructures containing tens of particles
\citep[for additional details see][]{Knebe2011}. Even when a halo
finder identifies substructures, their properties can be affected by
numerical resolution. Before accretion, the main effect of resolution
is on the inner structure of the subhalo. After accretion, poor
resolution can affect the orbit and tidal stripping of the
subhalo. While these effects are subdominant for subhalos resolved
with a large number of particles, they are very important for
subhalos resolved with around $100$ particles or less.

Resolution effects play an important role in establishing the extent
to which a given simulation can correctly probe the subhalo
population. In what follows we introduce a scaling method that allows
us to extrapolate the subhalo statistics beyond the resolution limit
of a simulation.  Applying this algorithm to an N-body simulation
involves two main steps:

\begin{enumerate}
\item[\textbf{I)}] Determining the range over which numerical effects
  influence the subhalo count. In general, a simulation correctly
  follows all substructures above a certain particle number, but
  resolves only a fraction of smaller subhalos. This
  results in missing substructures and a systematic underestimate of 
  the subhalo number count. \\[-.2cm]
\item[\textbf{II)}] Adding the missing subhalos in the range where
  only a partial subhalo population is found. This procedure recovers
  the mean and scatter of the subhalo abundance down to much lower
  subhalo masses than are resolved in the simulation. 
\end{enumerate}
In the remainder of this section we describe our method in more detail
and demonstrate how to use it to infer the true subhalo abundance in
the two Millennium Simulations.

\subsection{Step I: quantifying the resolution effects}
\label{subsec:quantifying_resolution_effects}

Since the CDM linear power spectrum of fluctuations has power on all
scales down to an Earth mass, ${\sim}10^{-6}\Msolar$, increasing the
resolution of a simulation results not only in a better determination
of the internal structure of high mass satellites, but also in the
generation of new, and previously not resolved, lower mass subhalos. To study finite resolution
effects, we consider the abundance of subhalos as a function of the
substructure to host size ratio. The mass of a subhalo is not a
well-defined quantity because it depends on the definition of the
subhalo's boundary and on the gravitational unbinding procedure.  A
more robust way to characterize subhalo size is through the maximum
circular velocity, $V_\rmn{max}$. This is determined by the inner
structure of the object and is therefore relatively insensitive to the
identification algorithm or the definition of boundary \citep[for
details see][]{Onions2012}. Furthermore, using $V_\rmn{max}$ to
characterise the size of satellites lends itself to a closer
comparison with observations that typically probe only the inner part
of a halo where the galaxy resides. Thus, rather than the mass ratio,
we will consider the ratio of $V_\rmn{max}$ to the host virial
velocity, $V_{200}$, defined as:
\begin{equation}
	V_{200} = \sqrt{ \frac{GM_{200}}{R_{200}} },
	\label{eq:v200}
\end{equation}
with $G$ the gravitational constant. 

We parametrise the substructure to host halo velocity ratio as
\begin{equation}
    \nu= \frac{V_\rmn{max}}{V_{200}} \;, 
\end{equation}
where $V_\rmn{max}$ refers to the subhalo and $V_{200}$ to the host
halo. We define $\Ncum$ as the average number of subhalos per host
with velocity ratio exceeding $\nu$. Given a sample of halos within
a chosen mass or $V_{200}$ range, the mean subhalo count is given by:
\begin{equation}
    \Ncum = \frac{ 1}{n_\rmn{hosts}} \sum_{i=1}^{n_\rmn{hosts}} N_i\cum{} \,,
\end{equation}
where $n_\rmn{hosts}$ denotes the numbers of halos in the sample and
$N_i\cum{}$ gives the number of subhalos with velocity ratio
exceeding $\nu$ in halo $i$. The derivative of this quantity, 
\begin{equation}
    \Nnu = \frac{d\Ncum}{d\nu}, 
\end{equation}
gives the mean number of subhalos per host with velocity ratio in the
range $\nu$ to $\nu+d\nu$ per $d\nu$ interval. 

Lack of numerical resolution will result in fewer than expected
substructures in an N-body simulation. For example, subhalos traced
by $\lsim100$ particles tend to have artificially low maximum circular
velocities because of the gravitational softening
\citep{aquarius2008}. The resulting lower concentration makes them
vulnerable to premature tidal disruption after they fall into the host
halo. We quantify the effects resolution on the subhalo number counts
by expressing, 
\begin{equation}
	\Nnu = \widetilde{N}(\nu) f(\nu), 
	\label{eq:completeness_function}
\end{equation}
where $\widetilde{N}(\nu)$ is the true subhalo count at $\nu$ in the
absence of resolution effects. The function $f(\nu)$ is the
completeness function that describes the artificial loss of subhalos
due to limited numerical resolution.  A value of $f(\nu)=1$ means that
the simulation has resolved all the substructures at $\nu$ while
values of $f(\nu)<1$ mean that only a partial population of subhalos
has been detected. Thus, quantifying this kind of resolution effect
reduces to measuring the completeness function, $f(\nu)$, for a given
simulation. 

\MCn{There is a wide range of factors that can influence the completeness function of cosmological simulations: gravitational softening length, integration timestep and other numerical parameters, to the halo finder and the code used to run the simulation. Exploring such a large parameter space to provide a general formula for $f(\nu)$ would be unfeasible, so instead we will show how to compute the function $f(\nu)$ for any given N-body simulation. Within the same simulation, the completeness function will likely depend on the mass of the host halo. We parametrise this dependence via the number of particles, $\mathcal{N}$,}
with which the host halo is resolved. Note that we use
$\overline{N}$ to denote the mean subhalo count and $\mathcal{N}$ to
denote the number of dark matter particles in the host halo.


\begin{figure}
     \centering
     \includegraphics[width=1\linewidth,angle=0]{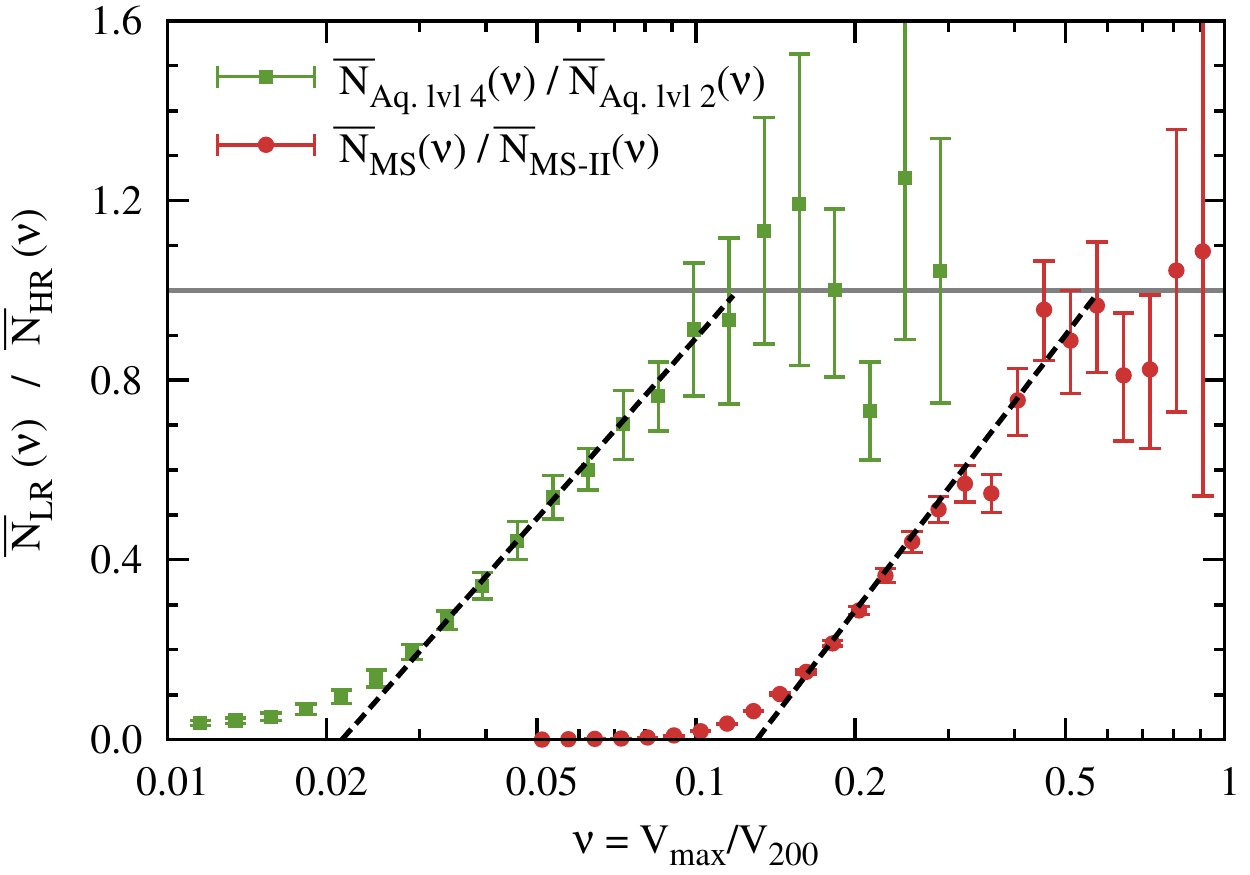}
     \caption{ \MCn{Comparison of the mean subhalo number, $\Nnu$, between equal mass haloes resolved at two different resolutions. The circles show halos in the mass range $(0.6 - 1.2)\times 10^{13}\Msolar$ that were resolved with $(0.7-1.4)\times 10^4$ particles in the \MI{} and with $125$ times more particles in the \MII{}. The squares compare $\Nnu$ of the Aquarius halos resolved with ${\sim}10^7$ particles at ``level-4'' and with ${\sim}20$ times more particles at ``level-2''.}
       The dashed curve shows that the
       transition from $1$ to $0$ is well approximated by a linear
       function in $\ln\nu$. The error bars represent the $1\sigma$
       uncertainty in the determination of the ratio between the two subhalo numbers. 
       }
     \label{fig:resolution_effects}
\end{figure}

To estimate the completeness function we compare the substructure
count between halos in simulations with two different
resolutions. The result is illustrated in
\reffig{fig:resolution_effects} where we contrast the mean subhalo
count of $(0.6 - 1.2)\times 10^{13}\Msolar$ mass halos that were resolved at low
resolution in \MI{} and at high resolution in the \MII{}. To emphasise the
difference we plot the ratio, $\overline{N}_\rmn{\MI}(\nu) /
\overline{N}_\rmn{\MII}(\nu)$, between the subhalo count in the two
simulations. Since ${\sim}10^{13}\Msolar$ mass halos in the \MII{} have
over $10^6$ particles, we expect $\overline{N}_\rmn{\MII}(\nu)$ to be
unaffected by numerical effects for $\nu\gsim0.15$ (for a detailed
justification of this point see
\refappendix{subsec:appendix_method_A}). This implies that for
$\nu\gsim0.15$ we have
$\overline{N}_\rmn{\MII}(\nu)\approx\widetilde{N}(\nu)$ and so,
according to \eq{eq:completeness_function}, the ratio
$\overline{N}_\rmn{\MI}(\nu) / \overline{N}_\rmn{\MII}(\nu)$ gives the
completeness function of \MI{} halos.

\reffig{fig:resolution_effects} shows that the completeness function
is flat and equal to 1 at values of $\nu>0.4$, indicating that
in that range the \MI{} recovers the full population of substructures. At
lower values of $\nu$, the completeness function decreases from 1 to 0
reflecting the fact that only a partial population of subhalos is
found in that range in the \MI{}. This is in agreement with the
qualitative expectation discussed above. The transition in the \MI{}
completeness function from 1 to 0 is well approximated by a linear
function of $\log\nu$, as shown by the dashed line in the
figure. Therefore, we can write the completeness function as:
\begin{equation}
	f(\nu) = 
    \begin{cases}
        1 & \nu\ge\nu_0  \vspace{.1cm} \\
        1 + \alpha \ln \left( \dfrac{\nu}{\nu_0} \right) & \nu_*<\nu<\nu_0   \vspace{.1cm} \\
        0 & \nu<\nu_*,
    \end{cases}
	\label{eq:fit_function}
\end{equation}
where $\alpha$ and $\nu_0$ are two free parameters (and $\ln$ denotes the natural logarithm). The $\alpha$
parameter gives the slope of the transition from 1 to 0, while $\nu_0$
gives the smallest value of $\nu$ for which the simulation identifies
all the substructures. The symbol, $\nu_*=\nu_0 e^{-1/\alpha}$,
denotes the point below which no more subhalos are detected. This
expression gives a very good match to the completeness function as
long as $f(\nu)\gsim0.2$, as can be seen in the figure. 

In \refappendix{subsec:appendix_method_B} we show that the two
parameter fit in \eq{eq:fit_function} gives a very good description of
the completeness function not only for the \MI{} and Aquarius haloes, but also for the
\MII{} and \dove{} simulations. Furthermore, we have checked that the
same holds true when using different halo finders.

\MCn{Thus, computing the completeness function of any given simulation reduces to finding the $\nu_0$ and $\alpha$ parameters introduced in \eq{eq:fit_function}.
We propose two different methods to calculate these parameters.} These procedures are described in detail in
\refappendix{sec:appendix_resolution} and can be summarised as
follows:
\begin{enumerate}
\item[$\bullet$] \textbf{Method A} is the standard procedure of
  comparing halos of equal mass in simulations of different
  resolution. We used this method to compute $f(\nu)$ for the \MI{} by
  comparing with the higher resolution \MII{} data. While this method
  is simple to implement, it has the drawback that it requires an
  additional simulation with ${\sim}100$ times higher mass resolution
  than the simulation of interest. Therefore, we can use method A for
  \MI{}, but not for the \MII{} and \dove{} since we do not have
  access to even higher resolution simulations. We introduce method A
  merely to show that our second technique, method B, gives reliable
  results.

\item[$\bullet$] \textbf{Method B} compares the subhalo population in
  low and high-mass halos in the same simulation. The procedure is
  based on the assumption that the mean subhalo count is self-similar
  amongst host halos of different masses \citep[see][and references
  therein]{Wang2012}. As we shall see in \refsec{sec:host_mass}, this
  assumption is satisfied to a good approximation for dark matter
  substructures but the addition of baryons and feedback processes
  would break the self-similar behaviour so it is unclear if this
  procedure can be modified to work in realistic hydrodynamical
  simulations of galaxy formation. Compared to method A, method B does
  not require a higher resolution simulation. This represents a great
  advantage and allows us to compute the completeness function for the
  \MII{} and \dove{} simulations.
\end{enumerate}

\begin{figure}
     \centering
      \includegraphics[width=1\linewidth,angle=0]{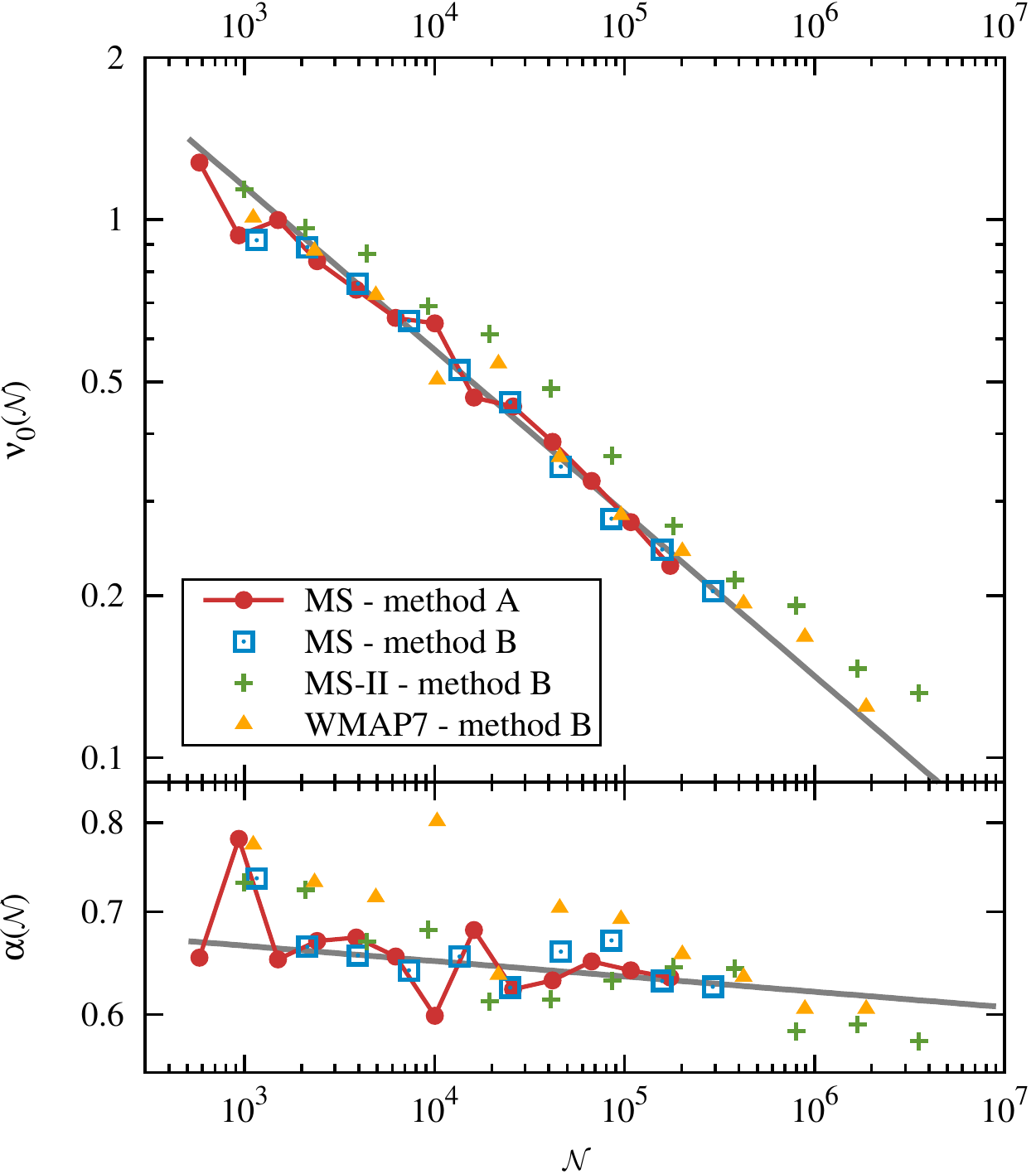}
      \caption{The dependence of the completeness function fit
        parameters, $\nu_0$ (top panel), and $\alpha$ (lower panel),
        on the number of particles, $\mathcal{N}$, in the host
        halo. The fit parameters were determined using the two
        different methods, A and B, described in
        \refappendix{sec:appendix_resolution}. The solid grey line
        shows a power-law fit to the results of method A. The power-law fits
        to each of the simulations are given in \reftab{tab:fit_parameters}. }
     \label{fig:fit_parameters}
\end{figure}

\begin{table}
  \caption{The values of the variables $\nu_0^0$, $n_{\nu_0}$, $\alpha^0$ and
    $n_{\alpha}$ given in \eq{eq:fit_parameters_N}. These
    quantities give the dependence of the fitting parameters of the
    completeness function, $\nu_0$ and $\alpha$, on the number of
    particles,  $\mathcal{N}$, in the host halo. We give values for subhalos
    within distance, $R_{200}$, and, $R_{100}$, from 
    the host halo centre. The $1\sigma$ error in the fit for 
    $\nu_0^0$ and $\alpha^0$ is 0.02, while that for 
    $n_{\nu_0}$ and $n_{\alpha}$  is 0.01.}
    \label{tab:fit_parameters}
    \begin{tabular}{lcccc}
        \hline
        Method - simulation & \parbox[][.9em][b]{0.1\linewidth}{\centering$\displaystyle\nu_0^0$} &  \parbox[][.9em][b]{0.1\linewidth}{\centering$\displaystyle n_{\nu_0}$} &  \parbox[][.9em][b]{0.1\linewidth}{\centering$\displaystyle \alpha^0$} &  \parbox[][.9em][b]{0.1\linewidth}{\centering$\displaystyle  n_{\alpha}$} \\
        \hline
        \multicolumn{5}{c}{$R_{200}$ substructures} \\
        \hline
        method A - \MI{}   &  0.57   & -0.30   & 0.65   & -0.01 \\
        method B - \MI{}   &  0.57   & -0.31   & 0.65   & -0.01 \\
        method B - \MII{}  &  0.67   & -0.29   & 0.67   & -0.02 \\
        method B - \dove{} &  0.57   & -0.29   & 0.72   & -0.03 \\
        \hline
        \multicolumn{5}{c}{$R_{100}$ substructures} \\
        \hline
        method A - \MI{}   &  0.55   & -0.30   & 0.65   & -0.02 \\
        method B - \MI{}   &  0.55   & -0.31   & 0.65   & -0.01 \\
        method B - \MII{}  &  0.67   & -0.29   & 0.65   & -0.03 \\
        method B - \dove{} &  0.56   & -0.28   & 0.72   & -0.04 \\
        \hline
    \end{tabular}
\end{table}

Using the two methods above we estimate the completeness function for
host halos of different mass. We find that the two fitting
parameters for $f(\nu)$ in \eq{eq:fit_function} depend most strongly
on the number of particles, $\mathcal{N}$, used to resolve the host
halo. This relationship is illustrated in
\reffig{fig:fit_parameters}. The $\alpha$ and $\nu_0$ parameters
show a power-law dependence on $\mathcal{N}$: 
\begin{equation}
    \nu_0(\mathcal N) = \nu_0^0 \left( \frac{\mathcal{N}}{10^4} \right)^{n_{\nu_0}}  \mbox{\,\,\, and \,\,\, }
    \alpha(\mathcal N) = \alpha^0 \left( \frac{\mathcal{N}}{10^4}
    \right)^{n_{\alpha}}. 
    \label{eq:fit_parameters_N}
\end{equation}
The quantities, $\nu_0^0$, $n_{\nu_0}$, $\alpha^0$ and $n_{\alpha}$,
are constants that depend on the numerical parameters of the
simulation, but not on $\mathcal{N}$. The two expressions in
\eq{eq:fit_parameters_N} give a very good description of
$\nu_0(\mathcal N)$ and $\alpha(\mathcal N)$. This is clearly shown in
the figure by the grey line which gives a power-law fit to the results
of method A applied to the \MI{} (solid red line with circular
symbols)\footnote{The power-law fits to $\alpha$ and $\nu_0$ shown in
  \reffig{fig:fit_parameters} work best for $\mathcal{N}\ge2000$. For
  halos resolved with fewer particles, the estimates of $\alpha$ and
  $\nu_0$ are less accurate due to the small number of points
  available for the fit. There is a degeneracy in the fit parameters
  $\alpha$ and $\nu_0$, since values with constant $\alpha\nu_0$ give
  similarly good fits. This introduces a large scatter in the two
  parameters around their mean trend with $\mathcal{N}$.}. The power
law fits to $\alpha$ and $\nu_0$ for the three simulations shown in
\reffig{fig:fit_parameters} are given in
\reftab{tab:fit_parameters}. All the simulations show the same
qualitative behaviour, though the exact values differ slightly. The
quantity $\nu_0$ varies as $\mathcal{N}^{-0.3}$, which is close to, but
shallower than the $\mathcal{N}^{-1/3}$ dependence that a naive
kinematic analysis would suggest. The parameter $\alpha$ varies only
slightly, as $\mathcal{N}^{-0.02}$.

\reffig{fig:fit_parameters} shows that the two methods, A and B, for estimating
the completeness function give the same results. This is clearly seen
when comparing the values of $\nu_0$ and $\alpha$ for the \MI{}
simulation obtained using method A (solid red line) and method B (blue
square symbols).  \MCn{Thus, $f(\nu)$ can be computed only using the information available in the simulation under study without the use of a higher resolution simulation, by following the method B procedure outlined in \refappendix{subsec:appendix_method_B}. }

\MCn{In addition, \reffig{fig:fit_parameters} shows that there are small differences between the completeness function of the three simulations studied here (see also \reftab{tab:fit_parameters}). Therefore, when precise results are needed, it is necessary to estimate $f(\nu)$ separately for each simulation. Computing the completeness function of a given simulation can be done with minimal computational resources using method B.}


The results presented up to now are for substructures within distance, 
$R_{200}$, of the host halo centre. We find that the fitting formula of
\eq{eq:fit_function} with very similar parameter values also describes
well the completeness function for subhalos within $R_{100}$ of the
host halo centre (see \reftab{tab:fit_parameters}).

\subsection{Step II: adding the missing subhalos}
\label{subsec:correcting_resolution_effects} 

As we have seen, the completeness function can be used to estimate the 
mean abundance of poorly resolved or unresolved subhalos as a function of
$V_\rmn{max}$. However, in practice, it is necessary to know 
not only the mean value of $\Ncum{}$, but also its dispersion $\sigma\cum{}$ 
across the halo population, which characterises the halo-to-halo
variation.

Given a completeness function, $f(\nu)$, lack of resolution implies
that a sample of $n_\rmn{hosts}$ halos are missing a fraction,
$1-f(\nu)$, of their substructures.  In total, the sample of halos is
missing a number of subhalos with velocity ratio, $\nu$, given by,
\begin{equation}
	n_\rmn{hosts} \; (1-f(\nu)) \; \widetilde{N}(\nu) = n_\rmn{hosts} \; \dfrac{1-f(\nu)}{f(\nu)} \; \Nnu{},
	\label{eq:missing_satellites}
\end{equation}
where $\widetilde{N}(\nu)$ and $\Nnu$ are the true and measured mean
subhalo count (\eq{eq:completeness_function}).  To recover the true
substructure count per halo, $\widetilde{N}(\nu)$, we add the missing
subhalos to the halo sample by randomly assigning each new subhalo to
a host. We take the probability that a new subhalo is assigned to host
halo, $i$, to be proportional to $1-f(\nu,\mathcal{N}_i)$, with
$\mathcal{N}_i$ the number of particles in host halo $i$. The special
case when the sample contains halos of similar mass corresponds to
each halo having equal weight, and so we distribute the missing
substructures among the hosts with equal probability.

We apply the procedure above to samples of halos within a narrow mass
range and repeat the process independently for samples of halos of
different mass. This assumes that halo mass is the only factor that
determines the subhalo count and ignores the effects of assembly
bias. Previous studies have show that the mean subhalo count depends
on halo properties other than mass, like concentration and formation
redshift \citep{Gao2004,Zentner2005a,Shaw2006,Gao2011}, as well as on
the large scale environment of the host
\citep{Busha2011,Cautun2013a}. Assembly bias can be taken into account
by further restricting the halo samples to hosts with a given
concentration or in a given environment. Neglecting assembly bias does
not affect the ability of the method to recover the true mean subhalo
count, but can result in a smaller value for the scatter in the
count. We do not expect this effect to be significant since
\cite{Gao2011} found that the dependence of the substructure number
count on halo properties is not the main driver of the observed
halo-to-halo scatter.

\begin{figure}
     \centering
     \includegraphics[width=1\linewidth,angle=0]{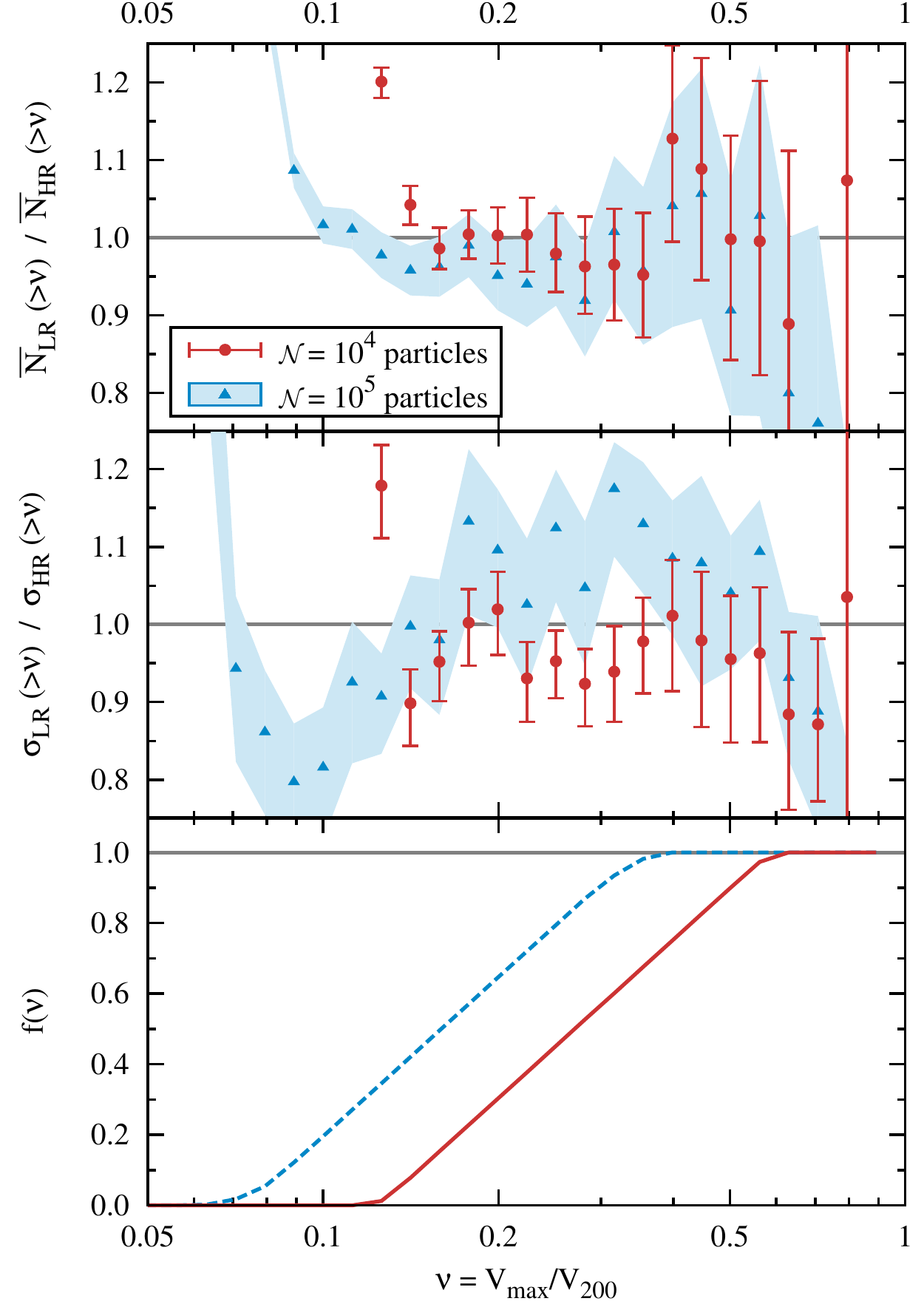}
     \caption{The effectiveness of our extrapolation method for
       subhalo statistics below the resolution limit of a
       simulation. The plots compare the mean, $\Ncum{}$ (top), and
       scatter, $\sigma\cum{}$ (middle) of the subhalo abundance in
       low and high resolution simulations. A value of one corresponds
       to a successful recovery of the mean and scatter. The low
       resolution data are \MI{} halos resolved with $(0.8-1.2)\times
       10^4$ (red circles) and $(0.4-1.2)\times 10^5$ (blue triangles)
       particles. The high resolution data are \MII{} halos of corresponding mass.
       The error bars show the $1\sigma$ uncertainty in the
       determination of $\Ncum{}$ and $\sigma\cum{}$. The bottom panel
       shows the completeness function, $f(\nu)$, corresponding to the
       low resolution halo samples: red for $\mathcal{N}{\sim}10^4$
       and dashed blue for $\mathcal{N}{\sim}10^5$. The extrapolation
       procedure is applied only in the region $f(\nu){<}1$; for
       $f(\nu){=}1$ there are no additional subhalos added. }
     \label{fig:cumulative_fit}
\end{figure}

\subsection{Evaluation of the extrapolation procedure}

\reffig{fig:cumulative_fit} shows how successful the extrapolation
method is in recovering the mean, $\Ncum{}$, and standard deviation,
$\sigma\cum{}$, of the subhalo population. The top panel gives the
ratio, $\overline{N}_\LR\cum /
\overline{N}_\HR\cum$, between the mean subhalo count found
at low and high resolution as a function of the velocity ratio,
$\nu$. The middle panel gives the ratio, $\sigma_\LR\cum /
\sigma_\HR\cum$, between the scatter in the subhalo counts
found at low and high resolution. In both cases a value of one
corresponds to a successful recovery of the true mean and scatter in
the number of subhalos. We illustrate the result of the extrapolation
method for host halos resolved with ${\sim}10^4$ (red circles) and
${\sim}10^5$ (blue triangles) particles in the \MI{} simulation. The two datasets
show the comparison for halos in the mass range $(0.69-1.1)\times10^{13}\Msolar$ 
and $(0.35-1.2)\times10^{14}\Msolar$ respectively, which were resolved at relatively 
low resolution in the \MI{} and at higher resolution in the \MII{}.

The bottom panel of \reffig{fig:cumulative_fit} shows the completeness
function, $f(\nu)$, of \MI{} halos resolved with ${\sim}10^4$ and
${\sim}10^5$ particles. For $f(\nu)=1$ there is no correction since
the number of new subhalos that need to be added is proportional to
$(1-f(\nu))/f(\nu)$ (see \eq{eq:missing_satellites}). The correction
becomes important only when $f(\nu)$ is significantly smaller than
unity. The top panel of the figure shows that we obtain
$\overline{N}_\LR\cum / \overline{N}_\HR\cum
\approx 1$ down to values of $\nu$ equal to 0.14 and 0.09 for halos
resolved with ${\sim}10^4$ and ${\sim}10^5$ particles
respectively. These values of $\nu$ correspond to the range where
$f(\nu)\gsim0.15$ as may be seen by comparing to the bottom panel of
the figure. Thus, our extrapolation method is successful at recovering
the true mean subhalo number count as long as $f(\nu)\gsim0.15$. 

\begin{figure}
     \centering
     \includegraphics[width=1\linewidth,angle=0]{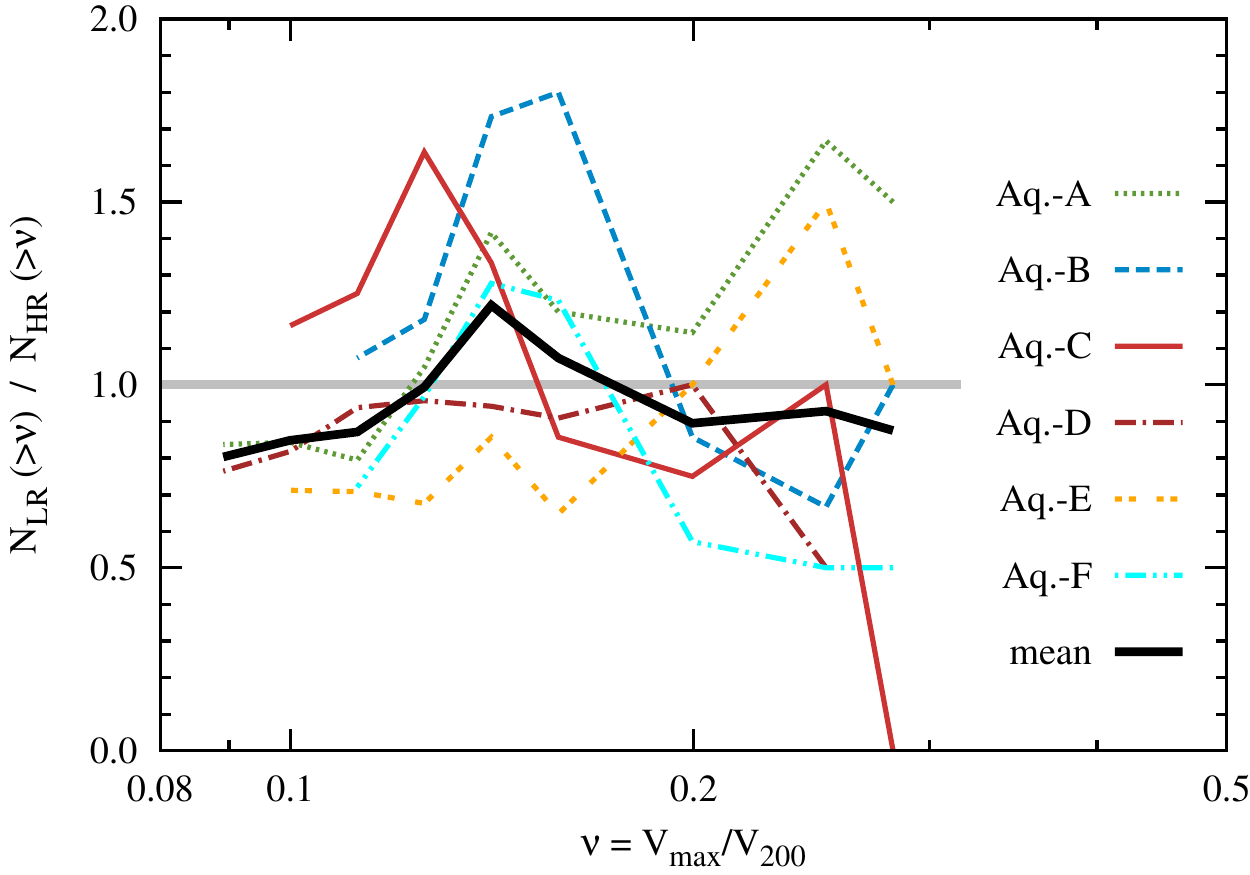}
     \caption{ An object-to-object comparison of the subhalo count for 
     the six Aquarius haloes, $N_\HR\cum$, and the corrected 
     subhalo count of their \MII{} counterparts, $N_\LR\cum$. The 
     solid black line compares the mean substructure number in the two 
     samples. The results reiterate that the extrapolation method gives the correct
     subhalo statistics although the scatter is appreciable for individual objects. }
     \label{fig:comparison_aquarius_m2}
\end{figure}

For the scatter in the subhalo number count, we find from the centre
panel of \reffig{fig:cumulative_fit} that $\sigma_\LR\cum /
\sigma_\HR\cum {\approx} 1$ down to values of $\nu$ of 0.16
and 0.11 for halos resolved with ${\sim}10^4$ and ${\sim}10^5$
particles respectively. Therefore, our extrapolation technique
recovers the correct subhalo scatter in the region where
$f(\nu)\gsim0.3$. In the case of the second dataset, we observe 
variations from unity of order $10\%$. These are due to the small sample of only $70$
\MII{} halos found in that mass range, which does not allow for a precise enough 
estimate of the scatter in the subhalo number count using bootstrap techniques. 
More importantly, we do not find any obvious
systematic effects in the estimate of $\sigma\cum{}$, except, at most,
a $5\%$ lower than expected value for $f(\nu)\lsim0.5$. This implies
that we can neglect subhalo assembly bias and still recover, to a good
approximation, the true subhalo scatter.  We also checked the
effectiveness of the extrapolation procedure for the \MII{} and
\dove{} simulations and found similar behaviour to the \MI{} case
presented here.

As a further test, we can compare the galactic mass halos in the
Aquarius simulations with their counterparts in the \MII{}
that have ${\sim}1000$ times fewer particles. With only six examples,
it is not possible to carry out a statistical comparison but since the
Aquarius halos are resimulations of \MII{} halos we can perform an
object-to-object comparison. This is shown in 
\reffig{fig:comparison_aquarius_m2}. We find that the ratio, 
$N_\LR\cum / N_\HR\cum$, between the low and high
resolution results oscillates around one, without any obvious systematic trend. 
This shows that the extrapolation method faithfully recovers the statistics of the
population over a large dynamical range in $\nu$.

\begin{figure}
     \centering
     \includegraphics[width=1\linewidth,angle=0]{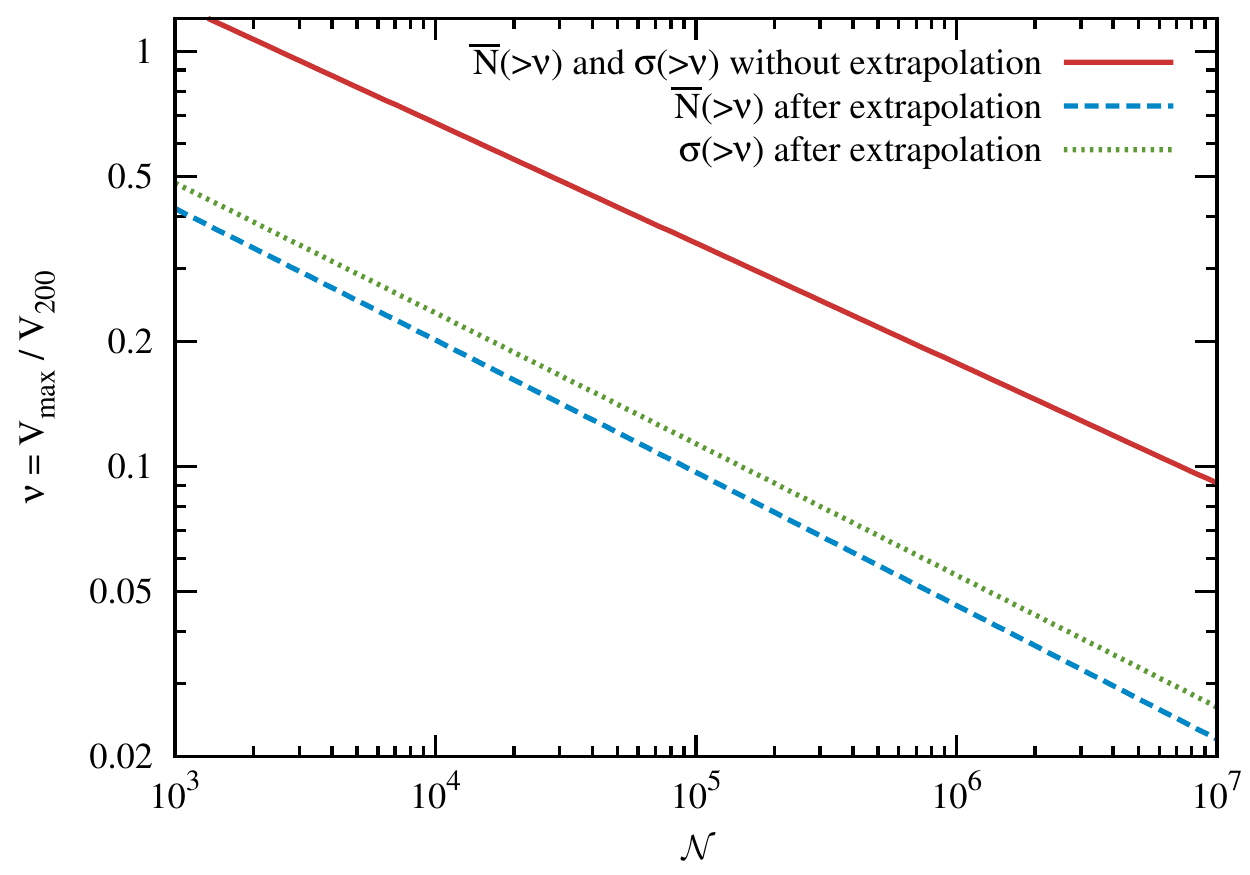}
     \caption{The lowest value of $\nu$ for which we recover the mean,
       $\Ncum{}$, and the dispersion, $\sigma\cum{}$, of the subhalo
       abundance. These limits are a function of the number of
       particles, $\mathcal{N}$, in the host halo. The solid curve
       gives the lower limits in the absence of extrapolation. The
       dashed and dotted curves give the of values of $\Ncum{}$ and
       $\sigma\cum{}$ respectively when extrapolating below the resolution
       limit. While the results shown here are for \MII{}, the other
       two simulations show a very similar behaviour, as may be seen in
       \reffig{fig:fit_parameters}. }
     \label{fig:expectation_corrected}
\end{figure}


In \reffig{fig:expectation_corrected} we show the values of $\nu$
above which we recover the true subhalo population with our
extrapolation method. The solid curve gives the $\nu$ limit in the
absence of extrapolation, given by the value of $\nu_0$ for \MII{}
from \reftab{tab:fit_parameters}. The dashed and dotted curves give
the $\nu$ limits for the mean and dispersion in the subhalo number
count when applying our extrapolation method. They were obtained by
solving the $f(\nu)=0.2$ and $f(\nu)=0.3$ equations and correspond to
conservative lower limits for recovering $\Ncum{}$ and $\sigma\cum{}$
as found in \reffig{fig:cumulative_fit}. By using our scaling method
we can estimate $\Ncum{}$ and $\sigma\cum{}$ to much lower $\nu$
values, corresponding to simulations with at least 50 times higher
mass resolution.


\section{The abundance of subhalos in MW-mass halos}
\label{sec:abundance}

\subsection{Mean subhalo number}
\label{subsec:abundance}

\begin{figure}
     \centering
     \includegraphics[width=1\linewidth,angle=0]{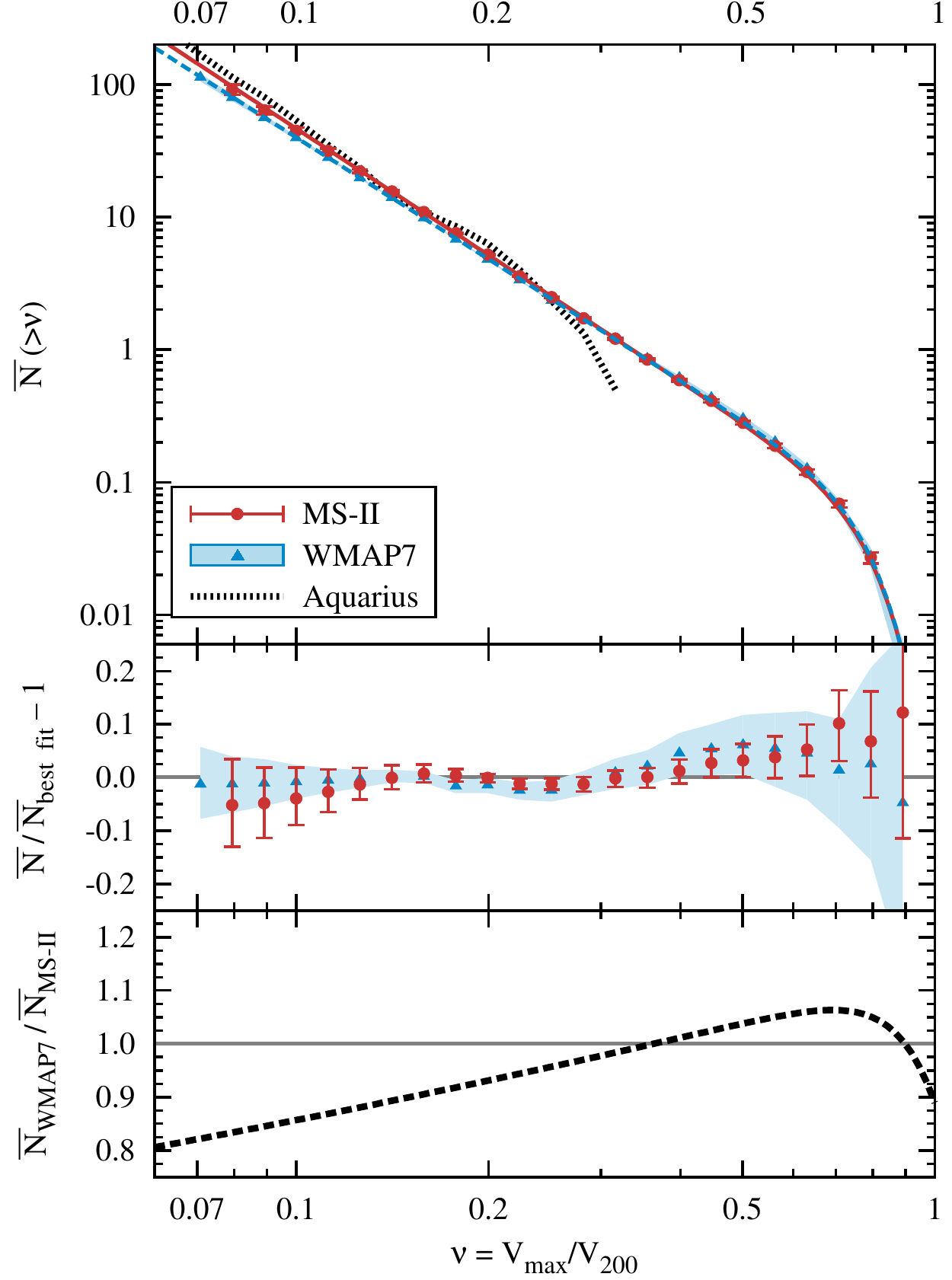}
     \caption{Top: the corrected mean subhalo count as a function of the
       velocity ratio, $\nu$, in MW-mass halos in the \MII{} and
       \dove{} simulations. The two lines show the best fit function
       given by \eq{eq:N_cumulative_best_fit} for \MII{} (solid red)
       and \dove{} (dashed blue). The dotted curve shows $\Ncum{}$ for
       the six Aquarius halos. Middle: the ratio between the actual
       number of subhalos in the simulations and the number given by
       the best fit function to the data in the top panel.  The
       errors bars give the $1\sigma$ error in the estimate of
       $\Ncum{}$ due to finite sample effects (which dominate for
       $\nu{>}0.3$) and due to uncertainties in the estimate of the
       completeness function (which dominate for $\nu{<}0.2$). Bottom:
       ratio of the best fit function to the subhalo abundance in the
       \MII{} and \dove{} simulations. }
     \label{fig:abundance_fit}
\end{figure}

In this section we investigate the subhalo distribution within halos
in the mass range $(0.6-2.2)\times10^{12}\Msolar$ to which we refer as
MW-like or MW-mass host halos. This mass range is consistent with
estimates of the MW halo mass obtained through a variety of methods
\citep{Wilkinson1999,Sakamoto2003,Battaglia2005,Dehnen2006,Smith2007,Li2008,Xue2008,Gnedin2010,Guo2010,Watkins2010,Busha2011b,Piffl2014}.

Using the extrapolation technique described in the previous section we
can recover the mean subhalo number, $\Ncum{}$, for $\nu\ge0.08$
(compared to $\nu\ge0.3$ in \MII{} and \dove{} in the absence of these
corrections). We illustrate this in \reffig{fig:abundance_fit} where
we show the corrected $\Ncum{}$ for MW-like hosts in the \MII{} and \dove{}
simulations. The mean
subhalo velocity function has a power law dependence at small $\nu$
and an exponential cut-off at large $\nu$. As \bkOne{} did, we find
that the function:
\begin{equation}
	N\cum{} = \left( \frac{\nu}{\nu_1} \right)^a  \exp\left( -\left( \frac{\nu}{\nu_\rmn{cut}} \right)^b \right)
	\label{eq:N_cumulative_best_fit}
\end{equation}
gives a good match to the cumulative mean number of substructures as a
function of $\nu$ for MW mass halos. Following the prescription given
by \bkOne{} we fit the mean subhalo abundance for both \MII{} and
\dove{}. The resulting best fit parameters for the two simulations are
given in \reftab{tab:N_best_fit}. The best fit function fits the data
very well, as may be seen in the middle panel of
\reffig{fig:abundance_fit}.

\begin{table}
  \caption{The best fit parameters of \eq{eq:N_cumulative_best_fit}
    for the mean subhalo number count in MW-like halos in the \MII{} and
    \dove{} simulations, within both $R_{200}$ and $R_{100}$ from the
    host halo 
    centre. The $1\sigma$ errors in the fit are at most, $\Delta a=0.02$,
    $\Delta \nu_1=0.003$, $\Delta b=1$ and $\Delta
    \nu_\rmn{cut}=0.02$. The value of the parameter $a$  is sensitive
    to errors in the estimate of the completeness function (see
    \reftab{tab:fit_parameters}), which introduces an additional systematic
    error of $\Delta a=0.07$. } 
    \label{tab:N_best_fit}
    \begin{tabular}{lccccc}
        \hline
        Simulation & Subhalos within & $a$ & $\nu_1$ & $b$ & $\nu_\rmn{cut}$ \\
        \hline
        \MII{}   & \multirow{2}{*}{$R_{200}$}    & -3.17  & 0.338  & 7  & 0.80  \\
        \dove{}  &                               & -3.05  & 0.336  & 7  & 0.79  \\
        \MII{}   & \multirow{2}{*}{$R_{100}$}    & -3.22  & 0.366  & 7  & 0.80  \\
        \dove{}  &                               & -3.12  & 0.364  & 7  & 0.79 \\
        \hline
    \end{tabular}
\end{table}

The \MII{} and \dove{} halos have the same number of massive
substructures, but there are important differences between the two
simulations for low values of $\nu$. The subhalo population in \MII{}
halos has a slightly steeper slope and thus a higher abundance at low
$\nu$ than in \dove{} halos.  From the bottom panel of
\reffig{fig:abundance_fit}, it can be seen that for WMAP-7
cosmological parameters, MW-like halos have only $93\%$ and $86\%$ of
the \MII{} subhalos at $\nu=0.2$ and $\nu=0.1$ respectively.

When comparing with results in the literature, we find that other
studies have systematically underestimated the substructure abundance
at low $\nu$ as a result of not taking finite resolution effects
properly into account. Thus, while \bkOne{} found similar values for
the $\nu_\rmn{cut}$ and $b$ fit parameters for \MII{} subhalos, they
underestimated the slope of the velocity function at low $\nu$: they
find $a=-2.98$ whereas for substructures within $R_{100}$ we find $a=-(3.22\pm0.09)$.
The discrepancy in slope is due to \bkOne{} fitting the subhalo
count down to $\nu=0.2$, while we find that without proper correction, 
the \MII{} simulation gives the correct subhalo abundance only for $\nu\ge0.3$.
In contrast, \wang{} found a
slope of $a=-3.11$ within $R_{200}$, which agrees within the errors
with our value of $a=-(3.17\pm0.09)$, but nevertheless they find
$20\%$ fewer subhalos within $R_{200}$ at all values of $\nu$. 
  
The main difference between \wang{} and \bkOne{} is that, just as we
have done, \wang{} used the invariance of $\Ncum{}$ with host halo
mass to estimate the average subhalo abundance. This approach appears
to give the correct value for the slope, $a$. However, as \bkOne{}
did, \wang{} overestimated the value of $\nu$ at which resolution
effects become important and their fits to $\Ncum{}$ included host
halos for which only ${\sim}75\%$ of substructures are detected.
Another difference with these studies is that we use a phase-space
halo finder while both \bkOne{} and \wang{} use a configuration-space
halo finder. However, we expect that this choice accounts for at most
a few percent of the difference\MCn{, as we show in \refappendix{sec:appendix_subfind}.}

\subsection{Scatter in the substructure population}
\label{subsec:scatter}
\begin{figure}
     \centering
     \includegraphics[width=1\linewidth,angle=0]{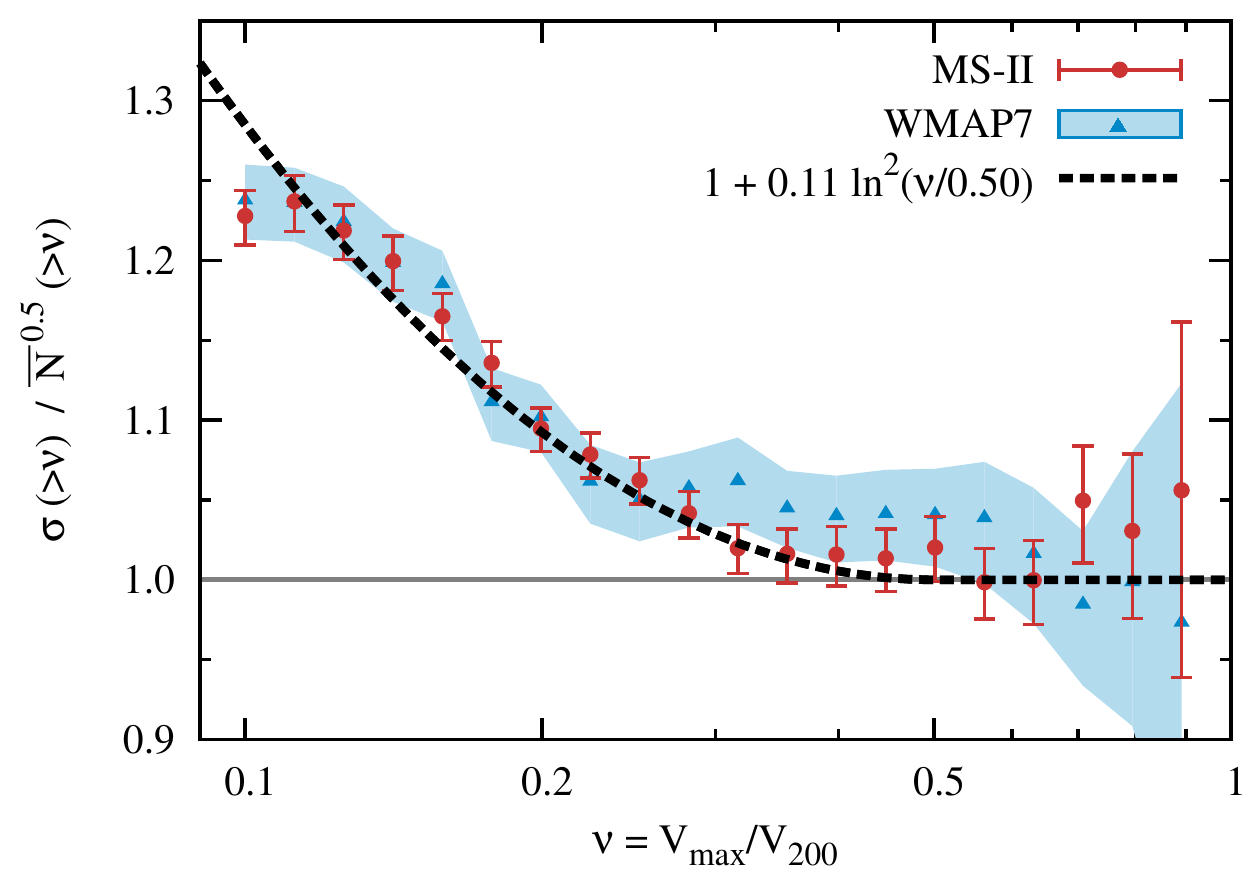}
     \caption{The dependence of the scatter in the subhalo abundance,
       $\sigma\cum{}$, on the velocity ratio, $\nu$, for MW-like
       hosts. For clarity we show
       $\sigma\cum{}/\overline{N}^{1/2}\cum{}$, the ratio between the
       observed scatter and the Poisson value,
       $\overline{N}^{1/2}\cum{}$. The dashed black curve gives the
       fit to the data for both the \MII{} and \dove{} simulations,
       with the best-fit parameters quoted in the legend. The error
       bars represent the $1\sigma$ error in $\sigma\cum{}$. }
     \label{fig:abundance_sigma}
\end{figure}

The dispersion of the subhalo number distribution characterises
halo-to-halo variations and is important not only for quantifying how
typical the MW and its satellites are, but also for the interpretation
of conclusions derived from very high-resolution simulations of a few
MW-sized halos \citep{Diemand2008,aquarius2008,Stadel2009}. The
scatter in the subhalo abundance is also an important parameter when
applying halo occupation distribution (HOD) models to populate dark
matter halos with galaxies
\citep[eg.][]{Benson2000,Seljak2000,Ma2000,Peacock2000,Scoccimarro2001,Berlind2002}. 

We find that, at large $\nu$, the scatter in the substructure
abundance matches the dispersion of a Poisson distribution with the
same mean. At lower velocity ratios, as the average number of
subhalos increases, we find a much larger scatter than expected for a
Poisson distribution. This is illustrated in
\reffig{fig:abundance_sigma} that shows the ratio of the measured
subhalo scatter to the dispersion, $\overline{N}^{1/2}\cum{}$, of a
Poisson distribution with mean $\Ncum{}$. We find that the standard
deviation, $\sigma\cum{}$, in both the \MII{} and \dove{} subhalo
distributions has the same dependence on $\nu$ that can be
parametrised as:
\begin{equation}
    \sigma\cum = \overline{N}^{1/2}\cum
    \begin{cases}
        1                         &  \nu\ge\nu_\sigma \\
        1 + \beta \ln^2(\nu/\nu_\sigma) &  \nu<\nu_\sigma.
    \end{cases}
    \label{eq:std_fit}
\end{equation}

Fitting this equation to the data, we find $\nu_\sigma=0.50$ and
$\beta=0.11$. This fit is a very good match to the subhalo abundance
scatter, as may be seen in \reffig{fig:abundance_sigma}. The scatter
for substructures within $R_{100}$ from the host halo centre shows a
similar functional form, but with best-fit parameters
$\nu_\sigma=0.55$ and $\beta=0.14$.

Our result that the scatter in the number of small substructures
differs significantly from the Poissonian expectation is in good
agreement with previous results: \cite{Benson2000} showed that
the occupation of halos by galaxies is not a Poisson process and \bkOne{}
showed that the dispersion in the number of subhalos above a certain
mass is a combination of Poisson scatter at the high mass end and
larger than Poisson scatter at the low mass end.

\subsection{Subhalo occupation distribution}
\label{subsec:distribution}
\begin{figure}
     \centering
     \includegraphics[width=1.\linewidth,angle=0]{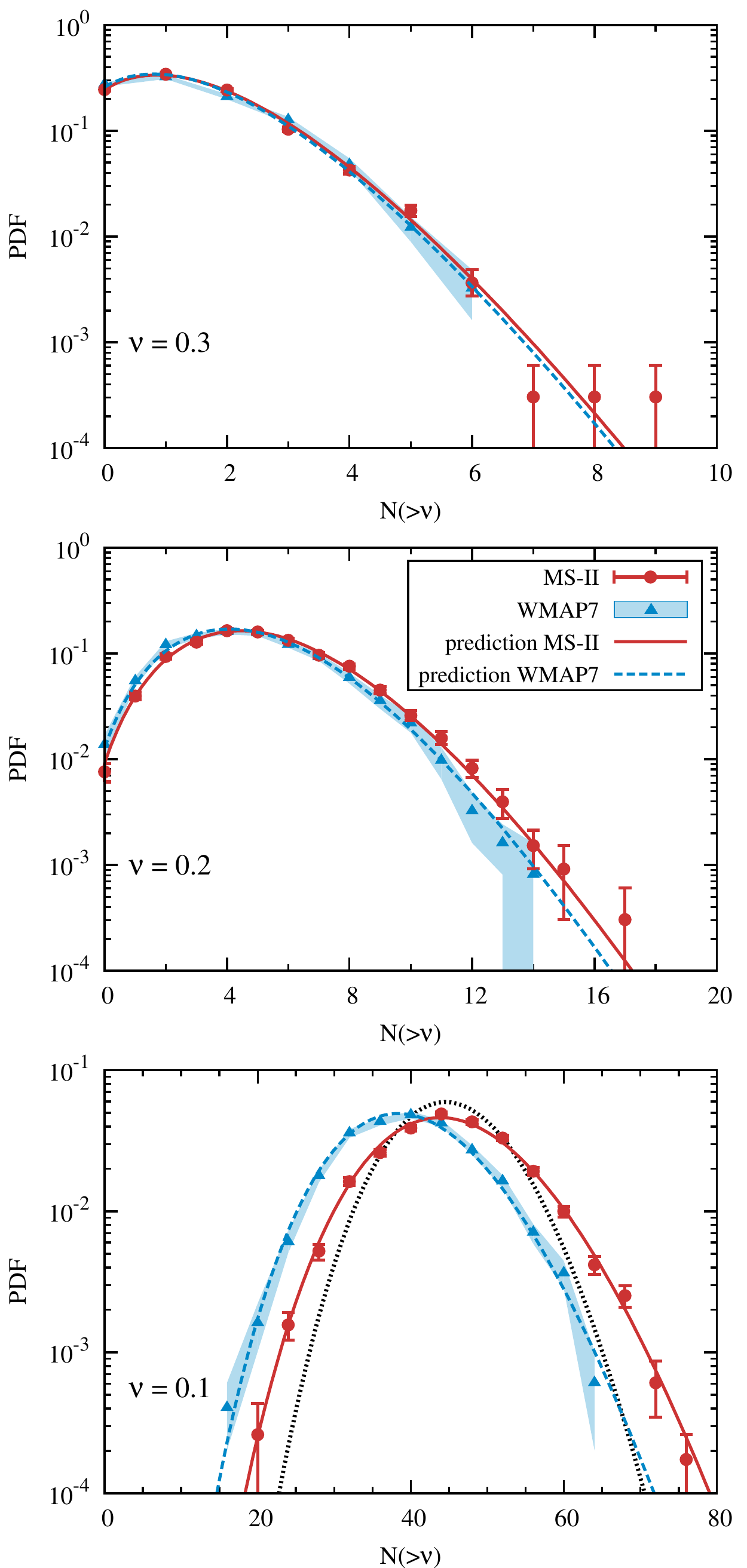}
     \caption{The probability distribution function of the
       number of substructures in MW-like host halos with
       $\nu\geq0.3$ (top), $\nu\geq0.2$ (middle) and 
       $\nu\geq0.1$ (bottom). The solid and dashed curves show
       the negative binomial distribution with the mean and standard
       deviation found in Figures~\ref{fig:abundance_fit} and
       \ref{fig:abundance_sigma}. The solid grey curve in the bottom
       panel shows a Poisson distribution with the same mean as the
       \MII{} halos.}
     \label{fig:distribution_all}
\end{figure}

Since the scatter in the subhalo population is significantly
non-Poissonian, we follow \bkOne{} and \cite{Busha2011}, and model the
probability distribution function (PDF) of the number of substructures
with a given value of $\nu$ using the negative binomial distribution
(NBD),
\begin{equation}
    P(N|r,s) = \frac{\Gamma(N+r)}{\Gamma(r)\Gamma(N+1)} s^r (1-s)^N,
    \label{eq:NBD}
\end{equation}
where $N$ is the number of subhalos per halo, $\Gamma(x)=(x-1)!$ is
the Gamma function, which, for integer values of $x$, reduces to the
factorial function, and $r$ and $s$ are two parameters. The mean and
dispersion of this distribution can be computed analytically in terms
of $r$ and $s$. The inverse holds too, with the two distribution
parameters given by:
\begin{equation}
    r = \frac{\mu^2}{\sigma^2-\mu} \mbox{ ,\hspace{1cm}  } s= \frac{\mu}{\sigma^2},
    \label{eq:NBD_parameters}
\end{equation}
where $\mu$ and $\sigma$ denote the mean and dispersion of the
NBD. Thus, $\mu$ and $\sigma$ completely specify the distribution.
The NBD has also been used to describe the number of satellite
galaxies in HOD models \citep[eg.][]{Berlind2002}.

\bkOne{} found that the NBD gives a better fit to the substructure
population than a Poisson distribution when counting all subhalos
containing more than a certain fraction of the host mass. We find that
the NBD also matches well the substructure PDF when counting all
subhalos with velocity ratios larger than $\nu$. This is illustrated
in \reffig{fig:distribution_all} where we plot the subhalo occupation
distribution for MW-mass hosts in both the \MII{} and \dove{}
simulations. The solid and dashed lines NBDs. These are not fits to
the data points, but are obtained from \eq{eq:NBD_parameters} using
the mean subhalo number, $\Ncum{}$, from \eq{eq:N_cumulative_best_fit}
and the dispersion, $\sigma\cum{}$, from \eq{eq:std_fit}. It is clear
in the figure that the NBD reproduces very well the subhalo
distribution at all values of $\nu$. Therefore, knowing the mean and
scatter of the subhalo number counts is enough to infer the full PDF.

The grey line in the lower panel of \reffig{fig:distribution_all}
shows a Poisson distribution with the same mean as the \MII{} subhalo
abundance. It is clear that the Poisson distribution severely
underestimates the tails of the PDF. Thus, even a modest increase in the
dispersion compared to the Poisson case ($25\%$ at $\nu=0.1$) leads to
large deviations from a Poisson distribution.


\section{Dependence of subhalo number on host mass}
\label{sec:host_mass}
\begin{figure}
     \centering
     \includegraphics[width=1\linewidth,angle=0]{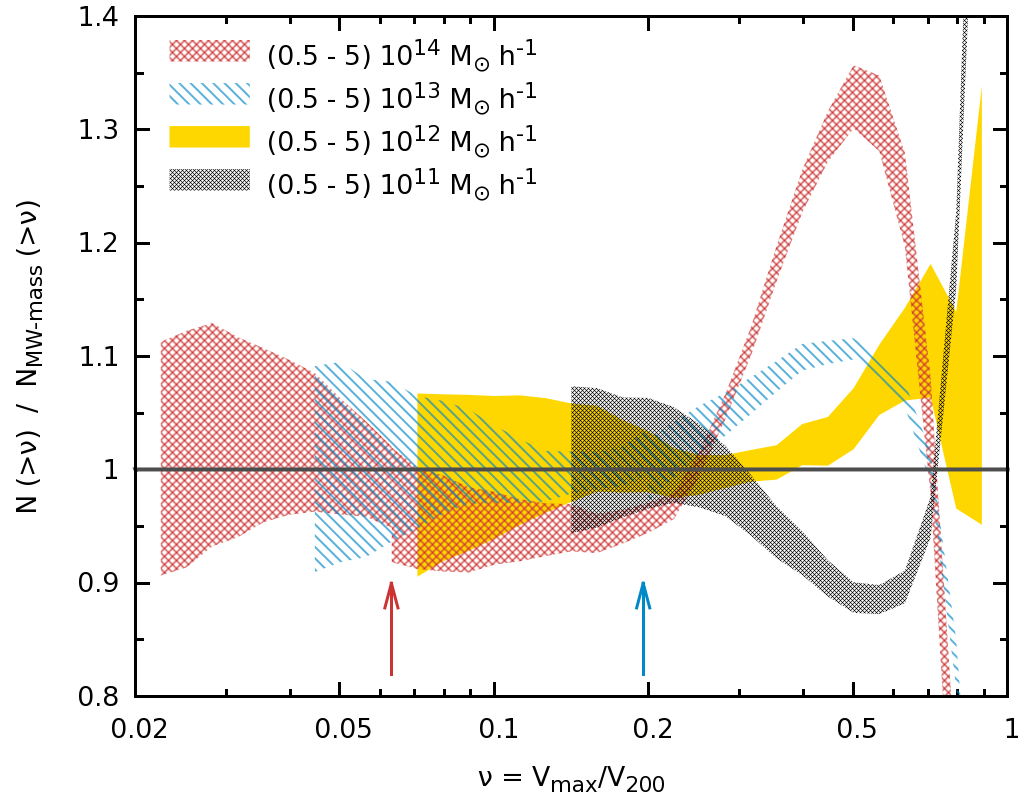}
     \caption{The dependence of the abundance of substructures as a
       function of $\nu$, $N\cum{}$, on the host halo mass. To
       emphasise the differences, we divide by the mean subhalo number,
       $\overline{N}_\rmn{MW-mass}\cum{}$, for \MII{} MW-like
       hosts. The two vertical arrows indicate the $\nu$ value where we
       switch from \MI{} to \MII{} data for hosts of mass
       $(0.5-5)\times10^{14}\Msolar$ (red arrow) and
       $(0.5-5)\times10^{13}\Msolar$ (blue arrow). The other two mass
       bins show only \MII{} halos. The width of each curve shows the 
       $1\sigma$ error in the determination of $N\cum{}$. }
     \label{fig:abundance_mass_variation}
\end{figure}

In \reffig{fig:abundance_mass_variation} we investigate how the mean
number of substructures as a function of normalized velocity, $\nu$,
varies for hosts of different mass. To emphasise the differences, we
normalise the mean subhalo number count in each mass bin by the mean,
$\overline{N}_\rmn{MW-mass}\cum{}$, for MW-mass hosts. We find that
for $\nu\le0.3$ there is very little dependence on host halo mass,
with at most a $5\%$ difference between MW-like and cluster sized
halos. In contrast, for larger subhalos we find a complex variation
with host mass that can be split in the two regimes.  Substructures with
$\nu\gsim0.8$ tend to be much more common in lower mass halos than in
high mass ones. Thus, it is much more likely to find a halo-subhalo
pair of similar mass in MW-like and lower mass hosts than in cluster
sized objects. In the $0.3\lsim\nu\lsim0.7$ range this trend is
reversed, with more subhalos present in massive hosts than in less
massive ones. In this case, the increase of $\Ncum{}$ with the mass of the host 
is small, with ${\sim}15\%$ variation in the number of substructures per 
decade of host halo mass.

The results in \reffig{fig:abundance_mass_variation} support the
assumption we made in \refsec{sec:resolution} that the mean subhalo
count as a function of $\nu$ varies only slowly or not at all with
host halo mass. This explains why method B works is valid for
estimating the completeness function. This also means that our results
for the subhalo population of MW-like halos are insensitive to the
exact mass range used to define MW-mass halos.  The invariance of the
mean substructure count on host mass makes it possible to use host
halos of all masses to compute $\Ncum{}$, but only for
$\nu\le0.3$. This property was already exploited by \wang{}, who used
halos in a large mass range to investigate the subhalo population of
MW-like halos and derive constraints on the MW halo mass.

The number of subhalos is independent of host halo mass only when
expressed in terms of the ratio
$\nu=V_\rmn{max}^\rmn{subhalo}/V_\rmn{200}^\rmn{host}$.
Previous studies have shown that there is a variation with host halo
mass when considering $\overline{N}({>}M^\rmn{subhalo}/M^\rmn{host})$
\citep{Gao2004,Zentner2005a,Gao2011} or
$\overline{N}({>}V_\rmn{max}^\rmn{subhalo}/V_\rmn{max}^\rmn{host})$
\citep{Klypin2011,Busha2011}.


\section{The MW massive satellites}
\label{sec:mw_probability}

As we discussed in \refsec{sec:introduction}, the conclusion that the
MW has at most three satellites residing in substructures with
$V_\rmn{max}\ge30\kms$ -- the two Magellanic Clouds and the
Sagittarius dwarf -- seems at odds with the number of such
substructures, eight on average, found in the Aquarius simulations of
halos of mass $M_{200} \sim 2\times 10^{12}\Msolar$
\citep{Boylan-Kolchin2011a}.  The probability of finding such a
population of substructures within \lcdm{} was investigated by \wang{}
who found that this ``too-big-to-fail-problem'' is only present if the
MW halo has a mass similar to the Aquarius halos, but the problem is
avoided altogether if the halo mass is a factor of 2 smaller. They
were therefore able to set an upper limit to the MW halo mass under
the assumption that \lcdm{} is the correct model. Since we find a
higher number of substructures than \wang{} did, we now re-examine how
their constraints on the MW mass change when using the subhalo
statistics derived in \refsec{sec:abundance}.

Given a halo of virial velocity, $V_{200}$, the probability that it
hosts at most $X$ substructures with $V_\rmn{max}\ge V_0$ is given by
\begin{equation}
    p({\le}X,V_0) = \sum_{k=0}^{X}  P(k|r\cum{},s\cum{}) \mbox{\hspace{.2cm} with} \;\nu=\frac{V_0}{V_{200}} \,,
    \label{eq:fraction_le}
\end{equation}
where $P(k|r\cum{},s\cum{})$ is the negative binomial distribution
that gives the probability that a halo has $k$ subhalos with
velocity ratio larger than $\nu$ (see Eq. \ref{eq:NBD}). The
distribution parameters, $r\cum{}$ and $s\cum{}$, are uniquely
determined by the mean and scatter of the subhalo population through
\eq{eq:NBD_parameters}.

\begin{figure}
     \centering
     \includegraphics[width=1\linewidth,angle=0]{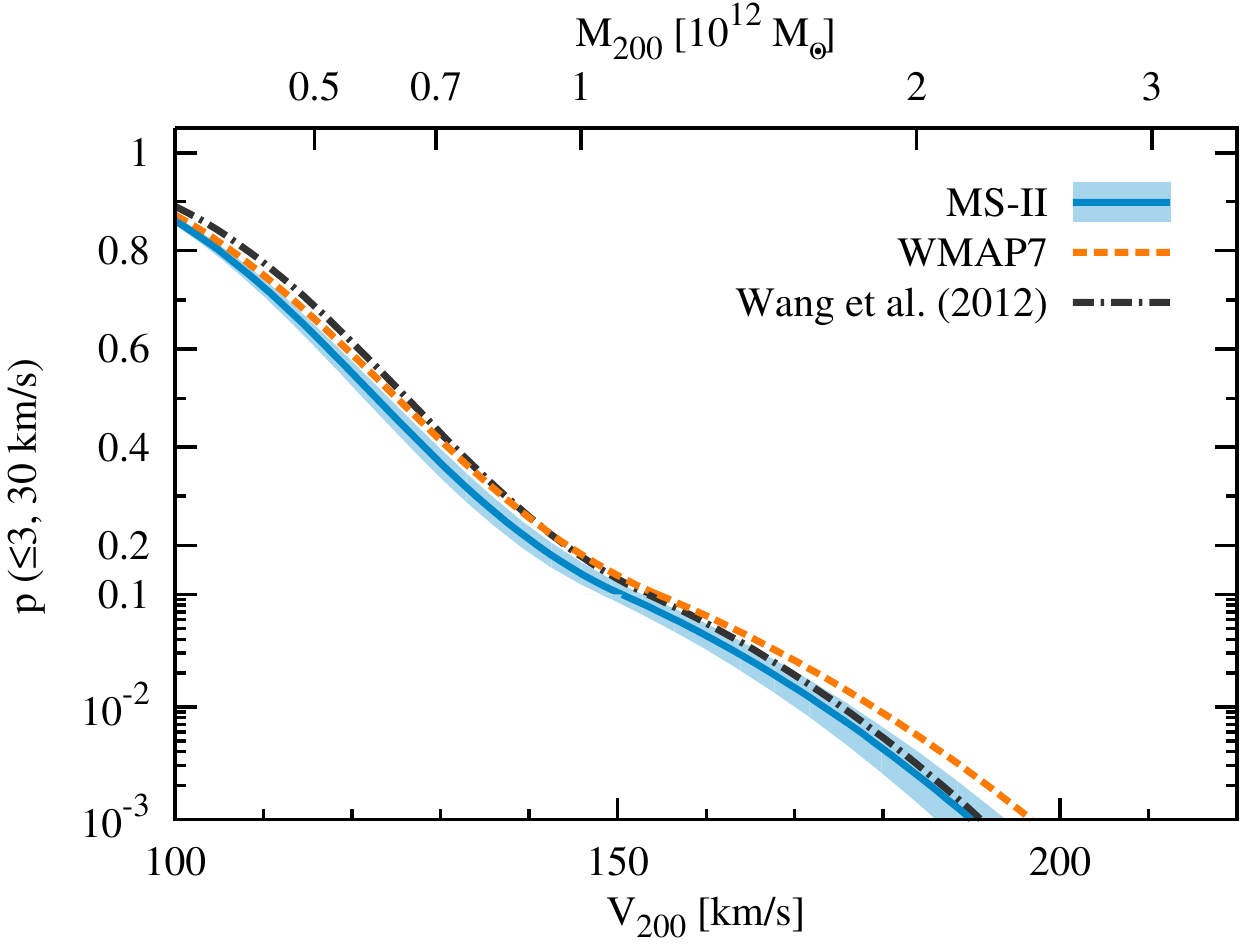}
     \caption{The probability, $p(\le3, 30\kms)$, that a halo has at
       most three substructures with $V_\rmn{max}\ge30\kms$ within a
       distance $R_{100}$ from its centre. The probability is given as
       a function of halo virial velocity, $V_{200}$ (lower tick
       marks), and halo mass, $M_{200}$ (upper tick marks). We show
       results for both the WMAP-1 cosmology used in \MII{} (solid
       line) as well as for the more recent WMAP-7 parameters (dashed
       line). The dashed-dotted line shows the results of \wang{}. 
       \MCn{The width of the \MII{} curve gives the $1\sigma$ error due to 
       uncertainties in the subhalo abundance of galactic halos. The WMAP7 results have the same error associated with them (not shown).}
       Note that the y-axis is linear above 0.1 and logarithmic for 
       lower values. }
     \label{fig:MW_constraint}
\end{figure}

The probability, $p(\le3, 30\kms)$, is shown in
\reffig{fig:MW_constraint} as a function of halo virial velocity
(lower tick marks) or, equivalently, halo mass (upper tick
marks). The results shown are for subhalos identified within $R_{100}$, which is
close to the maximum distance at which dwarf galaxies are identified
as being MW satellites. The solid blue curve shows $p(\le3, 30\kms)$
from the \MII{} simulation. The probability is a steep function of host
halo mass, decreasing from ${\sim}70\%$ at $0.5\times10^{12}\Msun$ to
${\sim}15\%$ at $1\times10^{12}\Msun$, and becomes negligible for
halos more massive than $2\times10^{12}\Msun$. For convenience, we
give values of $p(\le3, 30\kms)$ in \reftab{tab:MW_probability} for
some suggestive halo masses. Therefore, assuming that the \lcdm{}
cosmology is the correct model, given that the MW has only three
satellites with $V_\rmn{max}\ge30\kms$ it is unlikely that our
galaxy's halo is more massive than ${\sim}1.5\times10^{12}\Msun$.

The dashed orange curve in \reffig{fig:MW_constraint} shows results for
a \lcdm{} model with WMAP-7 parameters. Since this model has fewer
substructures at low $\nu$ than a model with WMAP-1 parameters, then,
at fixed halo mass, it has a higher $p(\le3, 30\kms)$ resulting in a
weaker upper limit on the MW halo mass.  Nevertheless, because of the
steep decline of the probability with halo mass, the upper limit on
the MW halo mass is only slightly increased in this model compared to
the one with WMAP-1 parameters.

\label{sec:MW_mass}

\begin{table}
    {\footnotesize
      \caption{The probability, $p({\le}3,30\kms)$, of finding three or
        fewer substructures with $V_\rmn{max}\ge30\kms$ for suggestive
        halo masses. Predictions are given for two choices of
        cosmological parameters derived from the 
        WMAP-1 and WMAP-7 data, and they are compared to the previous
        results 
        by \wang{} that are based on WMAP-1 parameters.} 
    \label{tab:MW_probability}
    \begin{tabular}{lccccc}
        \hline
        Halo mass & $[\times 10^{12}\Msun ]$ & \parbox[][][c]{0.034\textwidth}{\centering $0.5$} & \parbox[][][c]{0.034\textwidth}{\centering $0.7$}   & \parbox[][][c]{0.034\textwidth}{\centering $1$}   & \parbox[][][c]{0.034\textwidth}{\centering $2$}   \\
        \hline
        \parbox[][][c]{0.11\textwidth}{WMAP-1}          & $[\%]$ &  67 & 38  & 13 &  0.2 \\
        \parbox[][][c]{0.11\textwidth}{WMAP-7}          & $[\%]$ &  72 & 44  & 20 &  0.6 \\
        \parbox[][][c]{0.11\textwidth}{\wang{}}         & $[\%]$ &  76 & 48  & 22 &  0.3 \\
        \hline
    \end{tabular}
    }
\end{table}

\MCn{We expect that our results are robust to changes in cosmological parameters, especially concerning the recent \citet{Planck2013_XVI} measurement. The subhalo abundance could potentially be affected by the change in the concentration of halos and subhalos, but \citet{Dutton2014} showed that the increase in $\Omega_m$ between the WMAP-1 and Planck measurements is balanced by the decrease in the values of $\sigma_8$, $n_s$ and $h$, such that halos of the same mass have the same concentration in both cosmologies. In addition, increasing $\Omega_m$ leads to a larger number of halos at fixed mass and potentially to more subhalos, but, despite all this, the scaled subhalo velocity function is insensitive to variations in $\Omega_m$ \citep{Garrison-Kimmel2014}. }

Compared to the previous results of \wang{}, we find stricter upper
limits for the mass of the MW halo. This is seen in
\reffig{fig:MW_constraint} by comparing the solid and dashed-dotted
curves, with both corresponding to WMAP-1 parameters. The main cause of
the discrepancy is that \wang{} found up to $20\%$ fewer substructures
than we find (see \ref{subsec:abundance}) and thus overestimated the
probability at fixed halo mass. A second source of disagreement is the
PDF used to model the subhalo population. \wang{} used a Poisson
distribution that underestimates the true tails of the subhalo number
distribution (see \reffig{fig:distribution_all} for an example). This
effect becomes important when dealing with low $p(\le3, 30\kms)$
values and leads to an underestimate of the true probability. This is
the reason why the \wang{} probability for
$M_{200}\gsim10^{12}\Msun${} is lower than our value for WMAP-7
parameters, even though we find a larger subhalo count in the latter
case.


\section{How typical are the Aquarius halos?}
\label{sec:aquarius_halos}

In view of the prominence that the Aquarius halo simulations have had,
particularly in the work of \cite{aquarius2008} and
\cite{Boylan-Kolchin2011a,Boylan-Kolchin2012a}, it is interesting to
ask how typical these halos are of the global population of halos of
similar mass. \bkOne{} addressed this question in some detail using
the \MII{} and found that the six Aquarius halos are representative
in so far as the properties that they considered (such as assembly
history and internal structure) is concerned.  However, they did not
consider the distribution of $\nu$ that is of most interest here.

\begin{figure}
     \centering
     \includegraphics[width=1\linewidth,angle=0]{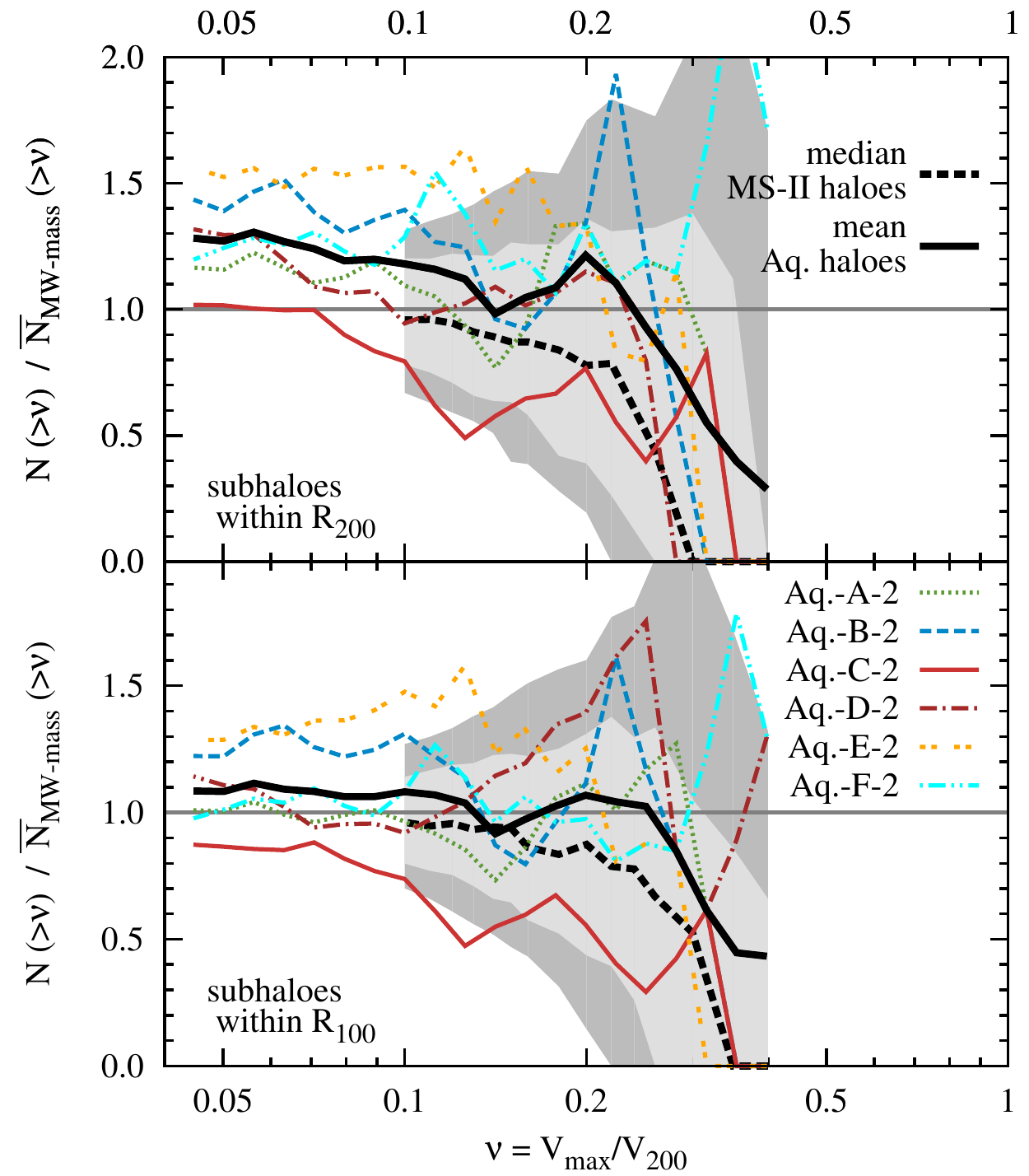}
     \caption{The cumulative distribution, $N\cum{}$, of normalized
       velocities in each of the six level-2 Aquarius halos compared
       with that of the global population of halos in the mass range 
       $(0.6-2.2)\times10^{12}\Msolar$ in the \MII{}. The Aquarius data
       are normalized to the mean, $\overline{N}_\rmn{MW-mass}\cum{}$,
       in the \MII{}. The top and bottom panels show results for
       substructures within distance $R_{200}$ and $R_{100}$ from the
       host halo centre respectively. The thick dashed curve shows the
       median of the \MII{} population while the light shaded region
       delimits the $16$ and $84$ percentiles of the distribution and
       the dark shaded region the $5$ and $95$ percentiles.  We
       restrict our analysis to $\nu\le0.4$, since only Aq.-F has
       subhalos with higher $\nu$. }
     \label{fig:aquarius_halos}
\end{figure}

In \reffig{fig:aquarius_halos} we compare the cumulative $\nu$
distribution, $N\cum{}$, of each of the six level-2 Aquarius halos
with that of the population of \MII{} halos in the mass range 
$(0.6-2.2)\times10^{12}\Msolar$. To show the differences more
clearly, we normalize the distributions to the the mean,
$\overline{N}_\rmn{MW-mass}\cum{}$, of the \MII{}.  The thick dashed
line shows the median for the \MII{} population, which is always
smaller than the mean count due to the long tail in the subhalo number
PDF (see \reffig{fig:distribution_all}). The light and dark shaded
regions show the $68\%$ and $90\%$ scatter around the median obtained
by modelling the subhalo distribution function as a NBD with the mean
and dispersion values given in \refsec{sec:abundance}.

Our comparison of the Aquarius and \MII{} subpopulations is restricted
to $\nu\ge0.1$, the resolution limit for \MII{} subhalos. For
completeness, we present the substructure function of the Aquarius
halos down to their resolution limit, $\nu\ge0.04$, but we do not use
the additional range in the comparison with the \MII{}.

\reffig{fig:aquarius_halos} shows that the six Aquarius halos have a
subhalo normalized velocity function that is in good agreement with
the much larger sample of \MII{} halos of similar mass, both when
considering subhalos within $R_{200}$ (top panel) and within $R_{100}$
(bottom panel). The mean of the Aquarius velocity function lies well
within the $68\%$ scatter and it is in very good agreement with the
mean in the \MII{}.  Individually, we find that Aq.-C has the
smallest number of subhalos compared to the other Aquarius halos,
especially for $\nu\le0.25$, but it is still within the \MII{}
distribution. The remaining five Aquarius halos have a substructure
velocity function similar or larger than the median for the \MII{}
halos. This result agrees with \wang{} who found that five of the
Aquarius halos have more substructures with $V_\rmn{max}\ge30\kms$
than the mean.

To summarize, for $\nu\lsim0.25$, five out of the six Aquarius halos
have a subhalo normalized velocity function that is similar or larger
than the median for a representative sample of MW-mass halos. This
needs to be born in mind when comparing the small number of massive
satellites present in the MW with the Aquarius halos, especially if,
as seems to be the case according to determinations of the satellite
luminosity function in the SDSS \citep{Guo2012,WWang2012}, our galaxy
has significantly fewer satellites than average.


\section{Summary}
\label{sec:conclusion}

We have introduced an extrapolation method to
infer subhalo number statistics below the resolution limit of a 
cosmological simulation. This method statistically generates the correct subhalo
abundance from the partial information available in a simulation of limited resolution. 
We have tested this technique by comparing
results of simulations of different resolution - including the high
resolution Aquarius simulations - and conclude that it extends the
subhalo number counts correctly down to what would be found in a
simulation of $50$ times or more better mass resolution. The technique
reproduces only the statistics of the subhalo population, not the
position or structure of subhalos.  We characterized the subhalo abundance
in terms of the scaled subhalo velocity function, $\Ncum{}$, which
gives the number of substructures above $\nu=V_\rmn{max}/V_{200}$,
where $V_\rmn{max}$ is the subhalo maximum circular velocity and
$V_{200}$ the virial velocity of the host.  We give a fitting formula
for the completeness function as a function of $\nu$ that can be used
to extrapolate the results of a simulation.

As noted by \bkOne{} and \cite{Busha2011}, the probability
distribution function of the number of substructures with a given
value of $\nu$ is well described by a negative binomial
distribution. Thus the substructure occupation distribution can be
obtained given only the mean and dispersion of the subhalo number
count. The scatter in the number counts becomes distinctly
non-Poissonian for $\nu\le0.3$, so simply assuming a Poisson
distribution will greatly underestimate the tails of the $\nu$ subhalo
distribution.

We applied our technique to the Millennium and Millennium-II (\MII{})
simulations and to a simulation of similar volume, but lower
resolution, with WMAP-7 cosmological parameters (rather than the
WMAP-1 values of the \MII{}).  We focused on halos of mass similar to
the MW but our results are insensitive to the exact halo
mass range assumed for the MW since the scaled subhalo velocity
function is insensitive to mass for $\nu\le0.3$; for larger values of
$\nu$ it shows a weak trend with host halo mass. This confirms, and
extends to a much larger dynamic range, the results of \wang{}
\citep[see
also][]{Moore1999,Kravtsov2004,Zheng2005,aquarius2008,Weinberg2008}.

As \bkOne{}, we found that the mean cumulative subhalo number count,
$\Ncum{}$, in halos of mass similar to the MW in the
\MII{} is well described by a power law with an exponential
cutoff. The number of small mass substructures depends slightly on the
cosmological parameters: it is lower for WMAP-7 than for WMAP-1
parameters,

We showed that the substructure population in halos of mass similar to
the MW in the \MII{} is complete only to $\nu{\sim}0.3$, which
corresponds to satellites with $V_\rmn{max}\sim45\kms$. By contrast,
our extrapolation method gives accurate results for the mean and
scatter of substructures in these \MII{} halos for $\nu\ge0.1$, which
corresponds to $V_\rmn{max}\sim15\kms$.  Previous studies
optimistically estimated that this subhalo population is complete down
to $\nu\sim0.2$.  \bkOne{} found ${\sim}15\%$ fewer subhalos than us
for $\nu\leq 0.2$. Exploiting the approximate scale-invariance of
$\Ncum{}$, \wang{} estimated the number of subhalos in MW-mass halos
over a large range of $\nu$.  However, they found $20\%$ fewer
substructures at all $\nu$ than we do because the $\nu$ function is
dominated by the low mass subhalos for which they recover only
${\sim}75\%$ of the population.


\wang{} used their inferred $\nu$ distribution of subhalos in MW-mass
halos and the fact that, as highligthed by 
\cite{Boylan-Kolchin2011a,Boylan-Kolchin2012a}, the MW has only a very
small number of massive satellites to set an upper limit on the MW
mass under the assumption that \lcdm{} is the correct cosmological
model.  Since we find fewer substructures in these halos
than \wang{} did, we revisited their argument and calculated the
probability for a halo to have a similar population of massive
substructure as the MW, i.e.  three or fewer substructures with
$V_\rmn{max}\ge30\kms$, as a function of the halo's mass. We were then
able to set a stricter upper bound on the MW mass than found by
\wang{}: the probability of having the observed number of large
subhalos is $20\%$ for $1\times10^{12}\Msun$ mass halos and
practically zero for halos more massive than $2\times10^{12}\Msun$.

Finally, we investigated how typical the subhalo population of the
Aquarius halos \citep{aquarius2008} is compared to those of the
global population of halos of similar mass in the \MII{}. We find
that the Aquarius halos fall within the scatter of the \MII{}
population but only one of the six Aquarius examples has fewer
subhalos than the median of the MW-mass halos in the \MII. This
needs to be born in mind when using the Aquarius subhalos to draw
general conclusions about our halo. 

\section*{Acknowledgements}

We are grateful to the referee's comments that have improved this paper.
This work was supported in part by ERC Advanced Investigator grant COSMIWAY 
[grant number GA 267291] and the Science and Technology Facilities Council 
[grant number ST/F001166/1, ST/I00162X/1].  WAH is also supported by the Polish
National Science Center [grant number DEC-2011/01/D/ST9/01960].
RvdW acknowledges support by the John Templeton Foundation, 
grant number FP05136-O. The simulations used in this study were carried out by the Virgo
consortium for cosmological simulations. Additional data
analysis was performed on the Cosma cluster at ICC in Durham and on the Gemini machines at the
Kapteyn Astronomical Institute in Groningen. 

This work used the DiRAC Data Centric system at Durham University, operated 
by ICC on behalf of the STFC DiRAC HPC Facility (www.dirac.ac.uk). This 
equipment was funded by BIS National E-infrastructure capital grant ST/K00042X/1, 
STFC capital grant ST/H008519/1, and STFC DiRAC Operations grant ST/K003267/1 and 
Durham University. DiRAC is part of the National E-Infrastructure.
This research was carried out with the support of the ``HPC Infrastructure for 
Grand Challenges of Science and Engineering'' Project, co-financed by the 
European Regional Development Fund under the Innovative Economy Operational Programme.


\newcommand{\jcap}{Journal of Cosmology and Astroparticle Physics}
\bibliographystyle{mn2e}
\bibliography{MW_subhaloes_bib}

\appendix

\section{Measuring the completeness function}
\label{sec:appendix_resolution}
We have employed two methods to investigate how the mean subhalo count
is affected by the finite resolution of an N-body simulation. In the
following we give a more detailed description of the two
methods, focusing on the advantages and limitations of each.

\subsection{Method A: comparing low and high resolution simulations}
\label{subsec:appendix_method_A}
\begin{figure}
     \centering
     \includegraphics[width=1\linewidth,angle=0]{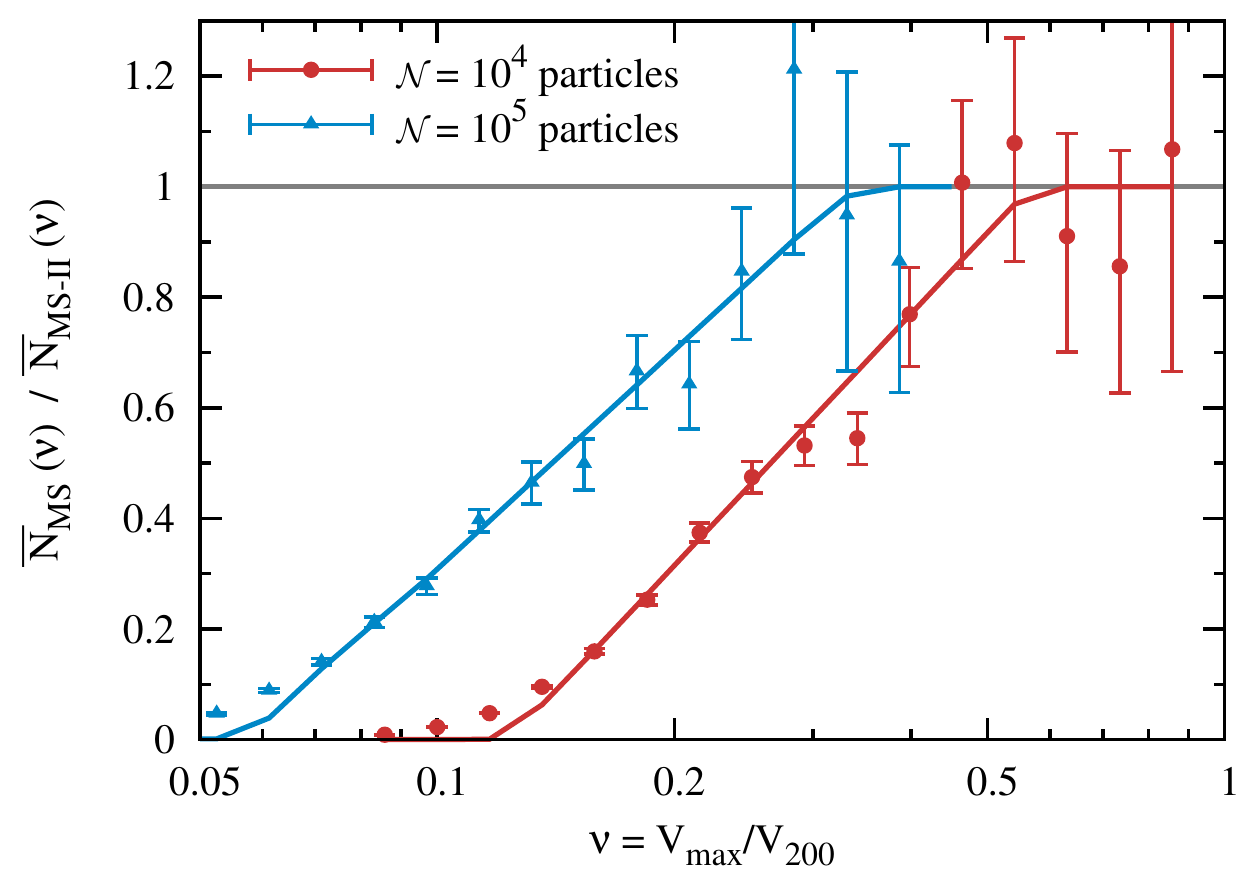}
     \caption{ Method A for computing the completeness function. This
       method compares the subhalo number count, $\Nnu$, in halos of
       a given mass resolved at two different resolutions in \MI{} and
       \MII{}. The two datasets consist of host halos in the mass
       range $(0.69 - 1.1)\times 10^{13}\Msolar$ (red filled circles)
       and $(0.6 - 1.2)\times 10^{14}\Msolar$ (blue filled
       triangles). These are resolved in \MI{} with $(0.8-1.2)\times
       10^4$ and $(0.7-1.3)\times 10^5$ particles respectively. The
       same halos are resolved in \MII{} with $125$ times more
       particles. The two solid lines represent the completeness
       function fit given by \eq{eq:fit_function}. The error bars
       represent the $1\sigma$ uncertainty in the determination of the
       $\overline{N}_\rmn{\MI}(\nu)/\overline{N}_\rmn{\MII}(\nu)$
       ratio.  
}
     \label{fig:method_A}
\end{figure}

The simplest way to investigate numerical effects is to compare the
subhalo population of halos of a given mass simulated at two different
resolutions. For this we use the two Millennium simulations
that resolve halos of similar mass with 125 times better resolution in
\MII{} than in \MI{}. Same mass halos have, on average, 
$\overline{N}_\rmn{\MI}(\nu)$ and $\overline{N}_\rmn{\MII}(\nu)$
substructures in \MI{} and \MII{} respectively. According to
\eq{eq:completeness_function}, the ratio of the two subhalo numbers is
given by  
\begin{equation}
	\frac{\overline{N}_\rmn{\MI}(\nu)}{\overline{N}_\rmn{\MII}(\nu)} = \frac{f_\rmn{\MI}(\nu)}{f_\rmn{\MII}(\nu)},
	\label{eq:different_resolutions}
\end{equation}
where $f_\rmn{\MI}(\nu)$ and $f_\rmn{\MII}(\nu)$ are the completeness
functions for the two Millennium simulations. Because of the higher
resolution of \MII{}, we can recover the full subhalo population down
to lower $\nu$ values than in \MI{}. Thus, this expression can be
rewritten as: 
\begin{equation}
	f_\rmn{\MI}(\nu) = \frac{\overline{N}_\rmn{\MI}(\nu)}{\overline{N}_\rmn{\MII}(\nu)},\quad \textrm{as long as}\,\, f_\rmn{\MII}(\nu)\cong 1.
	\label{eq:different_resolutions2}
\end{equation}
This holds down to the lowest value of $\nu$ for which \MII{} resolves
all substructures.

\reffig{fig:method_A} shows the ratio between the subhalo number
counts in the \MI{} and \MII{} simulations, for two samples of halos
in the mass range $(0.69 - 1.1)\times 10^{13}\Msolar$ and $(0.6 -
1.2)\times 10^{14}\Msolar$. The lower mass halos are resolved in
\MI{} with ${\sim}10^4$ particles while the higher mass ones are
resolved with ${\sim}10^5$ particles. We can see that resolution
effects become important at $\nu\approx0.6$ and $\nu\approx0.3$ for
halos resolved with $10^4$ and $10^5$ particles. By increasing the
number of particles by a factor of $10$, we would resolve the
subhalos down to approximately $2$ times lower values of $\nu$. Since
\MII{} has 125 times higher resolution than \MI{}, it recovers all
the subhalos down to ${\sim}4$ times lower $\nu$ than \MI{}, which,
according to the figure, corresponds to
$\overline{N}_\rmn{\MI}(\nu)/\overline{N}_\rmn{\MII}(\nu)\sim0.1$. This
means that we can use \eq{eq:different_resolutions2} to compute
$f_\rmn{\MI}(\nu)$ as long as $f_\rmn{\MI{}}(\nu)\gsim0.1$.

We find that the completeness function given by \eq{eq:fit_function}
gives a very good fit to the
$\overline{N}_\rmn{\MI}(\nu)/\overline{N}_\rmn{\MII}(\nu)$ ratio. This
is illustrated by the solid lines in \reffig{fig:method_A} for halos
resolved with $10^4$ and $10^5$ particles. The fit is a good match to
the completeness function for $\nu$ values for which $f(\nu)\ge0.2$. At
lower $\nu$ values the completeness function has a more complex
behaviour that is not captured by the two parameter expression that we
use. Therefore, we limit our analysis and fits to regions with
$f(\nu)\ge0.2$.

Method A for estimating the completeness function is very simple and
straightforward but its simplicity hides a major obvious disadvantage:
it requires a second simulation with ${\sim}100$ times higher mass
resolution than the original.  To overcome this limitation we
introduce a different method for computing the completeness function
which relies on a single simulation. We use method A to show that this
method B gives the same results.

\subsection{Method B: comparing low and high mass halos in the same simulation}
\label{subsec:appendix_method_B}

In a cosmological simulation subhalos are resolved to lower values of
$\nu$ in larger halos. Thus, if we assume that the true number of
subhalos as function of $\nu$, $\widetilde{N}(\nu)$ (see
\eq{eq:completeness_function}), is self-similar amongst host halos
of different mass (see \reffig{fig:abundance_mass_variation} and
\wang{}), then we can derive the completeness function by comparing
the substructure $\nu$ function in low versus high mass halos. 

\begin{figure}
     \centering
     \includegraphics[width=1\linewidth,angle=0]{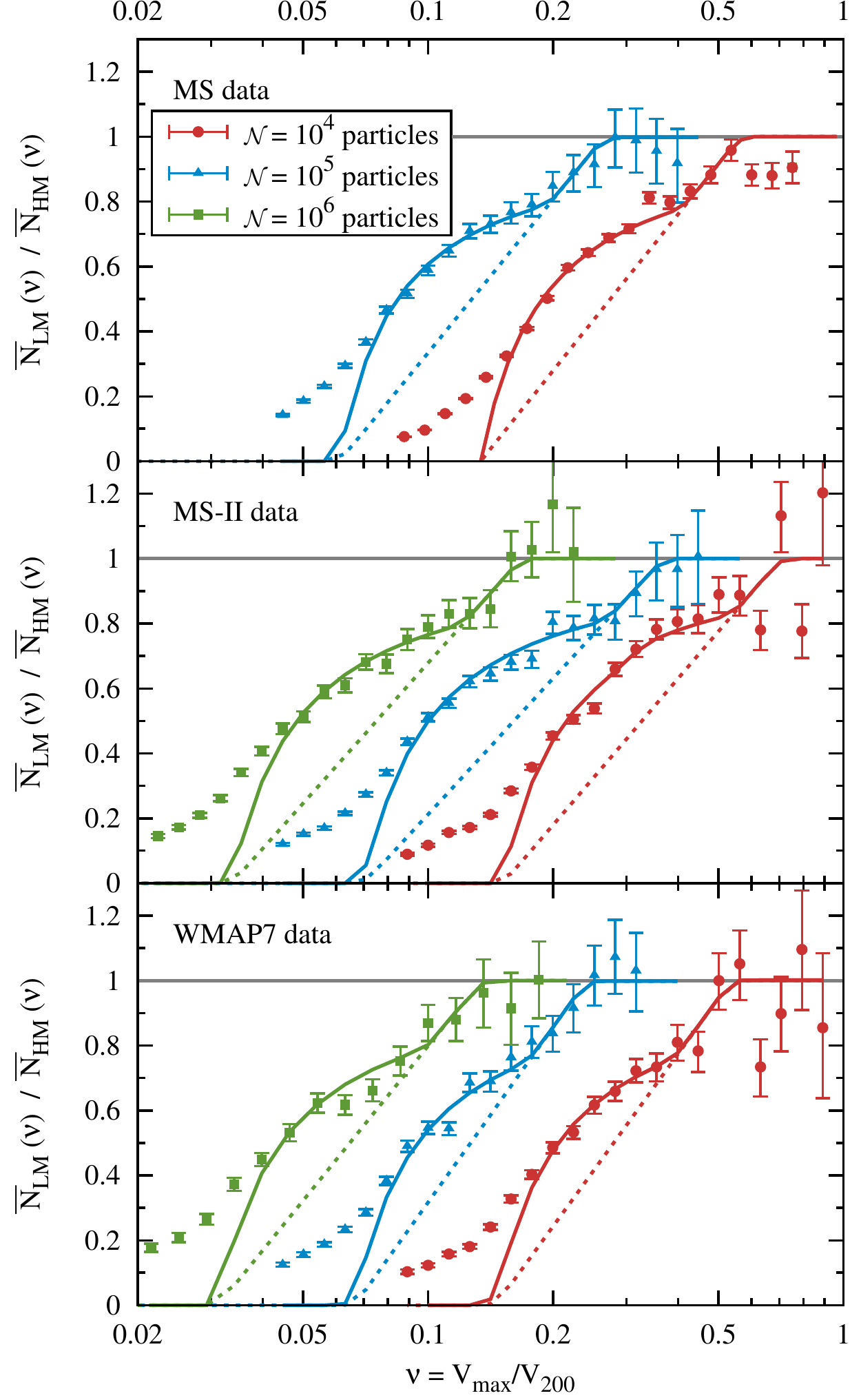}
     \caption{ Method B for computing the completeness function for
       the \MI{} (top panel), \MII{} (centre panel) and \dove{}
       (bottom panel) simulations. The method uses the ratio of the
       subhalo abundance, $\Nnu{}$, between low-mass (LM) and
       high-mass (HM) halo samples. The red, blue and green lines and
       symbols correspond to low-mass halos resolved with
       $(0.8-1.2)\times10^4$, $(0.8-1.2)\times10^5$ and
       $(0.8-1.2)\times10^6$ particles in each of the three
       simulations. The halo mass ranges used for each dataset are
       given in \reftab{tab:method_B_halo_mass}. The solid curves
       represent the fit given by \eq{eq:fit_function_different_mass}
       to each dataset. The dashed lines show the inferred
       completeness function, $f_\rmn{LM}(\nu)$, for the low-mass
       sample. The error bars represent the $1\sigma$ uncertainty in
       the determination of the
       $\overline{N}_\rmn{LM}(\nu)/\overline{N}_\rmn{HM}(\nu)$
       ratio. }
     \label{fig:method_B}
\end{figure}

\begin{table}
    \caption{ The mass range of halos resolved with
      $\mathcal{N}=(0.8-1.2)\times10^4$,
      $\mathcal{N}=(0.8-1.2)\times10^5$ and
      $\mathcal{N}=(0.8-1.2)\times10^6$ particles in the \MI{}, \MII{}
      and \dove{} simulations. These halo samples were used to obtain the
      results presented in \reffig{fig:method_B}. We only give the
      mass range for the low-mass halo sample, since halos in the
      high-mass sample are always $\Theta=3$ times more massive than these. } 
    \label{tab:method_B_halo_mass}
    \footnotesize{
    \begin{tabular}{lccc}
        \hline
        Simulation & $\mathcal{N}\sim10^4$ & $\mathcal{N}\sim10^5$ & $\mathcal{N}\sim10^6$ \\
        \hline
        \multirow{2}{*}{\MI{}}    & $(0.68-1.03)\times$ &  $(0.68-1.03)\times$ & -            \\
                                  & $10^{13}\Msolar$ &  $10^{14}\Msolar$ &                          \\[.2cm]
        \multirow{2}{*}{\MII{}}   & $(5.5-8.3)\times$   &  $(5.5-8.3)\times$   & $(5.5-8.3)\times$  \\
                                  & $10^{10}\Msolar$ &  $10^{11}\Msolar$ &  $10^{12}\Msolar$        \\[.2cm]
        \multirow{2}{*}{\dove{}}  & $(5.0-7.5)\times$   & $(5.0-7.5)\times$   & $(5.0-7.5)\times$   \\
                                  & $10^{10}\Msolar$ &  $10^{11}\Msolar$ &  $10^{12}\Msolar$  \\
        \hline
    \end{tabular}
    }
\end{table}

To illustrate this method we consider two halo samples: a low-mass
(LM) and a high-mass (HM) sample. Furthermore, we choose the high-mass
halos to be $\Theta$ times more massive than their low-mass
counterparts.  In the limit when $\widetilde{N}(\nu)$ is independent
of host halo mass\footnote{In reality $\widetilde{N}(\nu)$ varies
  slowly with host mass. To mitigate this effect we only compare halo
  samples that differ in mass only by a factor, $\Theta\sim$ a few.},
the ratio between the number of substructure in the low- and high-mass
samples is given by
\begin{equation}
	\frac{\overline{N}_\rmn{LM}(\nu)} {\overline{N}_\rmn{HM}(\nu)} = \frac{f_\rmn{LM}(\nu)} {f_\rmn{HM}(\nu)},
\end{equation}
where $f_\rmn{LM}(\nu)$ and $f_\rmn{HM}(\nu)$ are the completeness
functions of the two halo samples.  Using the $f(\nu)$ expression from
\eq{eq:fit_function}, the above relation becomes:
\begin{equation}
	\begin{cases}
        1 & \nu_0^\rmn{LM}\le\nu  \vspace{.1cm} \\
        1 + \alpha^\rmn{LM} \ln \left( \frac{\nu}{\nu_0^\rmn{LM}} \right) & \nu_0^\rmn{HM}\le\nu<\nu_0^\rmn{LM}   \vspace{.2cm} \\
        \dfrac{1+\alpha^\rmn{LM} \ln\left(\frac{\nu}{\nu_0^\rmn{LM}}\right)} {1+\alpha^\rmn{HM} \ln\left(\frac{\nu}{\nu_0^\rmn{HM}}\right)} & \nu_0^\rmn{LM} e^{-1/\alpha^\rmn{LM}}\le\nu<\nu_0^\rmn{HM}   \vspace{.2cm} \\
        0 & \nu<\nu_0^\rmn{LM} e^{-1/\alpha^\rmn{LM}}, 
    \end{cases}
	\label{eq:fit_function_different_mass}
\end{equation}
where $(\nu_0^\rmn{LM},\alpha^\rmn{LM})$ and
$(\nu_0^\rmn{HM},\alpha^\rmn{HM})$ are the completeness function fit
parameters corresponding to the low-mass and high-mass halo samples
respectively. This expression can be simplified further given 
that halos in the two samples are resolved with
$\mathcal{N}^\rmn{LM}$ and
$\mathcal{N}^\rmn{HM}=\Theta\mathcal{N}^\rmn{LM}$ particles. This,
combined with the dependence of the fit parameters, $\nu_0\propto
\mathcal{N}^{n_\nu}$ and $\alpha\propto \mathcal{N}^{n_\alpha}$, found
in \refsec{subsec:quantifying_resolution_effects}, results in
\begin{equation}
	\nu_0^\rmn{HM} = \nu_0^\rmn{LM}   \Theta^{n_\nu}    \mbox{~~~~and~~~~}
	\alpha^\rmn{HM} = \alpha^\rmn{LM} \Theta^{n_\alpha} \,.
	\label{eq:high_mass_params}
\end{equation}
Using these expressions reduces
\eq{eq:fit_function_different_mass} to 4 parameters: $\nu_0^\rmn{LM}$,
$\alpha^\rmn{ LM}$, $n_{\nu}$ and $n_\alpha$. These fit parameters can be found using the following algorithm:
\begin{list}{}{\leftmargin=3em \rightmargin=0em \labelwidth=3em}
    \item[(i)] Select a value for the mass ratio, $\Theta\sim$ a
      few\footnote{We have checked that the mass ratio, $\Theta$, 
        of the low mass and high mass samples does not affect 
        the fit parameters. While we recommend using $\Theta=3$, we have
        checked that similar fit parameters are obtained for
        $2\le\Theta\le10$. Using larger values of $\Theta$ introduces
        artifacts because of the mass dependence of $\Nnu$, while
        using smaller values results in very noisy fit parameters.}. 
    \item[(ii)] Make an initial guess for the parameters $n_{\nu}$ and $n_\alpha$.
    \item[(iii)] Select as the low-mass sample all halos in a chosen
      mass range. The high-mass sample then contains all halos
      $\Theta$ times more massive than this. Using these two samples
      find the best fit values of the parameters $\nu_0^\rmn{LM}$ and
      $\alpha^\rmn{LM}$.
    \item[(iv)] Repeat the previous step for different host halo
      masses in order to obtain the parameters $\nu_0^\rmn{LM}$ and
      $\alpha^\rmn{LM}$ for a wide range of halo masses. 
    \item[(v)] Use the dependence on mass, and therefore on host
      particle number, $\mathcal{N}$, of $\nu_0^\rmn{LM}$ and
      $\alpha^\rmn{LM}$ found in the previous step to find new values
      for $n_\nu$ and $n_\alpha$. 
    \item[(vi)] Check if $n_\nu$ and $n_\alpha$ have converged to the
      values used as the input for step \textit{(iii)}. If the values have
      converged, stop the iterative procedure. Otherwise, repeat steps
      \textit{(iii)} through \textit{(vi)} using the latest values for
      $n_\nu$ and $n_\alpha$.
\end{list}

In \reffig{fig:method_B} we illustrate the use of method B to compute
the completeness function for the three N-body simulations used in
this study. The figure shows the ratio,
$\overline{N}_\rmn{LM}(\nu)/\overline{N}_\rmn{HM}(\nu)$, of the
mean number of subhalos in the low- and high-mass halo samples. We
plot this ratio for low-mass halos resolved with ${\sim}10^4$,
${\sim}10^5$ and ${\sim}10^6$ particles, with masses given in
\reftab{tab:method_B_halo_mass}. To minimize the variation of the
subhalo number counts with mass we take the high-mass sample to be
$\Theta=3$ times more massive than the low-mass one. The fit given by
\eq{eq:fit_function_different_mass} is shown as a solid curve for each
of the datasets. We can see that it gives a very good fit for
$\overline{N}_\rmn{LM}(\nu)/\overline{N}_\rmn{HM}(\nu)\geq0.4$, which
corresponds to values $f_\rmn{LM}(\nu)\geq0.2$, the same limit 
for which Method~A is also accurate. 

\reffig{fig:method_B} shows another important result. The completeness
function has the same parametric form, given by \eq{eq:fit_function},
for all the three simulations used in this study. This is a reflection
of the fact that \eq{eq:fit_function_different_mass} gives a very good
fit to the $\overline{N}_\rmn{LM}(\nu)/\overline{N}_\rmn{HM}(\nu)$
ratio for the three simulations: \MI{}, \MII{} and \dove{}.

Computing the completeness function using method B has the advantage
of not requiring a simulation with a higher mass resolution as in
method A. This opens up the possibility of quantifying how numerical
effects in any given simulation alter the mean subhalo abundance. We
illustrated this for \MII{} and \dove{} for which we do not have a
higher resolution version and so we cannot apply method~A.  The main
limitation of method~B stems from the assumption that the mean subhalo
abundance is self-similar amongst host halos of different mass. As we
found in \refsec{sec:host_mass}, this condition is satisfied for
substructures in dark matter only simulations, but it will not be the
case when adding in baryons. The complex feedback processes involved
in galaxy formation affect halos of different mass in different ways
\citep[eg.][and references within]{Sawala2013}. This breaks the
self-similar behaviour of the subhalo abundance.

\section{Comparison of \textsc{rockstar} and \textsc{subfind} subhalo abundances}
\label{sec:appendix_subfind}

\begin{figure}
     \centering
      \includegraphics[width=\linewidth,angle=0]{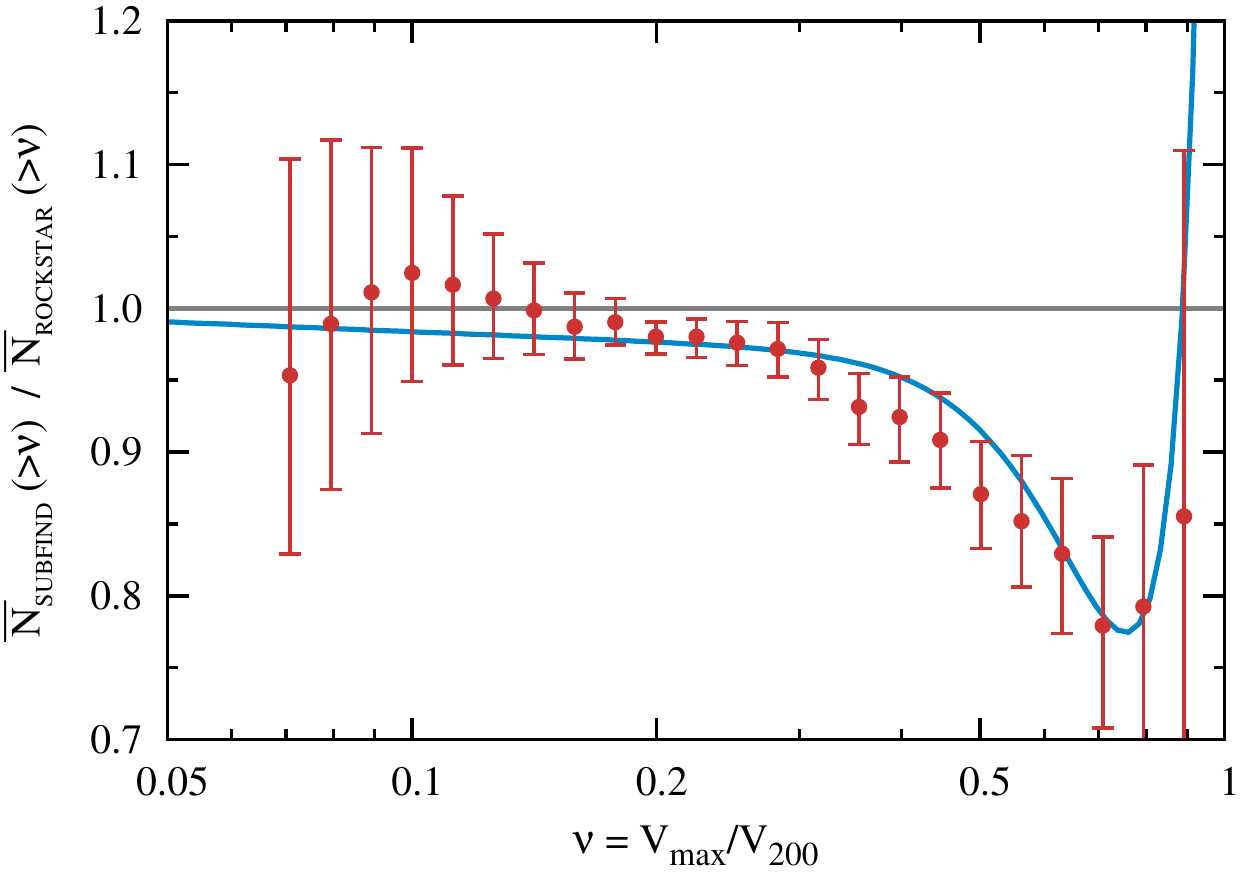}
      \caption{ \MCn{Comparison of the subhalo abundance of galactic mass halos identified with \subfind{} versus that found with \rockstar{}. The points give the $\overline{N}_\rmn{\subfind}(\g\nu)/\overline{N}_\rmn{\rockstar}(\g\nu)$ ratio as measured in the \MII{}. The solid curve shows the ratio between the best fit function (see \eq{eq:N_cumulative_best_fit}) to the \subfind{} and \rockstar{} subhalo abundance.} }
     \label{fig:comparison_Subfind_Rockstar}
\end{figure}

\MCn{Here we investigate if the difference in the subhalo numbers between our analysis and previous studies can be explained by the use of different halo finders. For this, we compare the galactic subhalo abundance as found by \rockstar{} \citep{Berhoozi2011} and by \subfind{} \citep{Springel2001b}, with the latter used in the studies of \wang{} and \bkOne{}. }

\MCn{We apply the same analysis steps to \subfind{} subhalos as we did in the case of \rockstar{}: identify the number of missing substructures due to resolution effects and estimate the true subhalo abundance, following the procedure described in \refsec{sec:resolution}. The resulting subhalo abundance for MW-mass hosts is well described by \eq{eq:N_cumulative_best_fit} with best fit parameters: $a=-3.18$, $\nu_1=0.333$, $b=6$ and $\nu_\rmn{cut}=0.78$ (for subhalos found within a distance $R_{200}$ from the host). \reffig{fig:comparison_Subfind_Rockstar} compares the subhalo abundance found with \rockstar{} and \subfind{}, showing that for $\nu\lsim0.3$ both halo finders get the same number of substructures, up to a few percent difference. For higher $\nu$, \subfind{} identifies ${\sim}10\%$ fewer substructures. Given that such massive subhalos are resolved with $\gsim10^3$ particles, the difference is likely due to substructures found close to the centre of the host that are identified by \rockstar{}, which is a phase-space halo finder, and not by \subfind{}, which uses only real-space information. Since \wang{} computed the subhalo abundance only in the interval $0.1\le\nu\le0.5$, the figure clearly shows that the use of \rockstar{} instead of \subfind{} cannot on its own explain the ${\sim}20\%$ higher subhalo abundance found in our study. Similarly, the significantly lower value of the subhalo abundance slope, $a$, found by \bkOne{} is not due to the use of a different halo finder. }

\end{document}